\documentclass[twocolumn]{aastex62}
\usepackage{natbib}

\newcommand{\angstrom}{\textup{\angstrom}}

\graphicspath{{./}{figures_submit/}}

\received{2019 April 15}
\revised{2019 June 4}
\accepted{2019 June 12}
\submitjournal{ApJ}

\shorttitle{BH Mass Measurements of $z\sim6$ Low-L Quasars}
\shortauthors{Onoue et al.}

\begin{document}

\title{Subaru High-$z$ Exploration of Low-Luminosity Quasars (SHELLQs). VI. Black Hole Mass Measurements of Six Quasars at $6.1\leq z \leq 6.7$}

\author[0000-0003-2984-6803]{Masafusa Onoue}
\email{onoue@mpia-hd.mpg.de}
\affiliation{Max Planck Institut f\"ur Astronomie, K\"onigstuhl 17, D-69117 Heidelberg, Germany}
\affiliation{National Astronomical Observatory of Japan, 
2-21-1, Osawa, Mitaka, Tokyo 181-8588, Japan}
\affiliation{Department of Astronomical Science, 
Graduate University for Advanced Studies (SOKENDAI),
 2-21-1, Osawa, Mitaka, Tokyo 181-8588, Japan}

 \author{Nobunari Kashikawa}
 \affil{Department of Astronomy, School of Science, The University of Tokyo,
 Tokyo 113-0033, Japan.}
\affiliation{National Astronomical Observatory of Japan, 
2-21-1, Osawa, Mitaka, Tokyo 181-8588, Japan}
\affiliation{Department of Astronomical Science, 
Graduate University for Advanced Studies (SOKENDAI),
 2-21-1, Osawa, Mitaka, Tokyo 181-8588, Japan}

\author{Yoshiki Matsuoka}
\affiliation{Research Center for Space and Cosmic Evolution, Ehime University, 
Matsuyama, Ehime 790-8577, Japan.}

\author{Nanako Kato}
\affiliation{Research Center for Space and Cosmic Evolution, Ehime University, 
Matsuyama, Ehime 790-8577, Japan.}

\author{Takuma Izumi}
\affiliation{National Astronomical Observatory of Japan, 
2-21-1, Osawa, Mitaka, Tokyo 181-8588, Japan}

\author{Tohru Nagao}
\affiliation{Research Center for Space and Cosmic Evolution, Ehime University, 
Matsuyama, Ehime 790-8577, Japan.}

\author{Michael A. Strauss}
\affiliation{Princeton University Observatory, Peyton Hall,
Princeton, NJ 08544, USA.}

\author{Yuichi Harikane}
\affiliation{National Astronomical Observatory of Japan, 
2-21-1, Osawa, Mitaka, Tokyo 181-8588, Japan}
\affiliation{Institute for Cosmic Ray Research, The University of Tokyo, 5-1-5 Kashiwanoha, Kashiwa, Chiba 277-8582, Japan}
\affiliation{Department of Physics, Graduate School of Science, The University of Tokyo, 7-3-1 Hongo, Bunkyo, Tokyo, 113-0033, Japan}

\author{Masatoshi Imanishi}
\affiliation{National Astronomical Observatory of Japan, 
2-21-1, Osawa, Mitaka, Tokyo 181-8588, Japan}
\affiliation{Department of Astronomical Science, 
Graduate University for Advanced Studies (SOKENDAI),
 2-21-1, Osawa, Mitaka, Tokyo 181-8588, Japan}
 
\author{Kei Ito}
\affiliation{National Astronomical Observatory of Japan, 
2-21-1, Osawa, Mitaka, Tokyo 181-8588, Japan}
\affiliation{Department of Astronomical Science, 
Graduate University for Advanced Studies (SOKENDAI),
 2-21-1, Osawa, Mitaka, Tokyo 181-8588, Japan}
 
\author{Kazushi Iwasawa}
\affiliation{ICREA and Institut de Ci\`encies del Cosmos, Universitat de Barcelona, 
IEEC-UB, Mart\\CID{3}ii Franqu\`es, 1, 08028 Barcelona, Spain.}

\author{Toshihiro Kawaguchi}
\affiliation{Department of Economics, Management and Information Science, Onomichi City University, 
Onomichi, Hiroshima 722-8506, Japan.}

\author{Chien-Hsiu Lee}
\affiliation{National Optical Astronomy Observatory, 950 North Cherry Avenue, 
Tucson, AZ 85719, USA.}

\author{Akatoki Noboriguchi}
\affiliation{Research Center for Space and Cosmic Evolution, Ehime University, 
Matsuyama, Ehime 790-8577, Japan.}

\author{Hyewon Suh}
\affiliation{Subaru Telescope, National Astronomical Observatory of Japan (NAOJ), National Institutes of Natural Sciences (NINS), 650 North A'ohoku place, Hilo, HI 96720, USA}

\author{Masayuki Tanaka}
\affiliation{National Astronomical Observatory of Japan, 
2-21-1, Osawa, Mitaka, Tokyo 181-8588, Japan}
\affiliation{Department of Astronomical Science, 
Graduate University for Advanced Studies (SOKENDAI),
 2-21-1, Osawa, Mitaka, Tokyo 181-8588, Japan}

\author{Yoshiki Toba}
\affiliation{Department of Astronomy, Kyoto University, Kitashirakawa-Oiwake-cho, Sakyo-ku, Kyoto 606-8502, Japan.}
\affiliation{Academia Sinica Institute of Astronomy and Astrophysics, 11F of Astronomy-Mathematics Building, AS/NTU, No.1, Section 4, Roosevelt Road, Taipei 10617, Taiwan.}
\affiliation{Research Center for Space and Cosmic Evolution, Ehime University, 
Matsuyama, Ehime 790-8577, Japan.}





\begin{abstract}
We present deep near-infrared spectroscopy of six quasars at $6.1\leq z \leq 6.7$ with VLT/X-Shooter and Gemini-N/GNIRS.
Our objects, originally discovered through a wide-field optical survey with the Hyper Suprime-Cam (HSC) Subaru Strategic Program (HSC-SSP), have the lowest luminosities ($-25.5$ mag $\leq M_\mathrm{1450}\leq-23.1$ mag) of the $z>5.8$ quasars with measured black hole masses.
From single-epoch mass measurements based on Mg{\sc ii} $\lambda2798$, 
we find a wide range in black hole masses,
from $M_\mathrm{BH}=10^{7.6}$ to $10^{9.3} M_\odot$.
The Eddington ratios $L_\mathrm{bol}/L_\mathrm{Edd}$ range from $0.16$ to $1.1$, but the majority of the HSC quasars are powered by $M_\mathrm{BH}\sim10^9M_\odot$ SMBHs accreting at sub-Eddington rates.
The Eddington ratio distribution of the HSC quasars is inclined to lower accretion rates 
than those of 
\citet{Willott10a}, who measured the black hole masses for similarly faint $z\sim6$ quasars.
This suggests that the global Eddington ratio distribution is wider than has previously been thought.
The presence of $M_\mathrm{BH}\sim10^9M_\odot$ SMBHs at $z\sim6$ cannot be explained with constant sub-Eddington accretion from stellar remnant seed black holes.
Therefore, we may be witnessing the first buildup of the most massive black holes in the first billion years of the universe, 
the accretion activity of which is transforming from active growth to a quiescent phase.
Measurements of a larger complete sample of $z\gtrsim6$ low-luminosity quasars, as well as deeper observations with future facilities will enable us to better understand the early SMBH growth in the reionization epoch.

\end{abstract}



\section{Introduction} \label{sec:Sec1}
Quasars are among the most luminous objects in the universe, and are powered by mass accretion onto supermassive black holes (SMBHs).
Since the 2000s, wide-field optical and near-infrared (NIR) surveys have discovered more than $200$ quasars at $z>5.7$ \citep[e.g.,][]{Fan01, Mortlock11, Venemans13, Reed15, Jiang16, Banados16, Wang17, Mazzucchelli17, Banados18, Wang18,  Fan19}.
One of the most remarkable aspects of the high-redshift quasars is that they are typically powered by SMBHs more massive than one billion solar masses
\citep[e.g.,][]{Mortlock11, Wu15, Banados18}, 
comparable to the most massive black holes at any redshift.
For these gigantic black holes, the growth timescale from their seeds is so short that they would need to have near-constant Eddington-limit accretion from the Big Bang, if they originated from the remnants of Population-{\sc III} (Pop-{\sc III}) stars ($M_\mathrm{seed}\lesssim10^{2-3} M_\odot$; \citealt{Heger03, Hirano14}).
This fact has put strong constraints on scenarios for SMBH formation and early growth history.
The most popular scenarios to explain the mass assembly of several billion solar-mass black holes in the early universe include intense gas accretion in a super-Eddington phase \citep[e.g.,][]{Kawaguchi04, Pezzulli16, Inayoshi16}, and direct collapse of primordial gas clouds resulting in $10^{5-6}M_\odot$ seed black holes \citep[e.g.,][]{Loeb94, Latif13, Chon16}.
Recent reviews on seed black formation models can be found in \citet{Volonteri10} and \citet{Latif16}.

Measurements of black hole masses of high-redshift quasars allow us to determine their Eddington ratios, gaining insight into their accretion history.
As the identification of $z\gtrsim6$ quasars is based on observations of  Ly$\alpha$ emission, independent follow-up observations are required to trace other broad emission lines that can be used to estimate the virial black hole mass. 
In particular, Mg{\sc ii} $\lambda2798$ is the line most $z\gtrsim6$ mass measurements rely on, as Balmer lines such as H$\alpha$ and H$\beta$ fall beyond the NIR coverage of ground-based spectrographs.
The C{\sc iv} $\lambda1549$ emission line is another frequently used mass estimator, but it is known to be affected by nuclear-scale outflows \citep[e.g.,][]{Shen16, Coatman17, Shen19}.
At $z>5.8$, $60$ quasars to date have Mg{\sc ii}-based mass measurements \citep{Kurk07, Jiang07, Willott10a, Mortlock11, DeRosa11, DeRosa14, Wu15, Mazzucchelli17, Banados18, Eilers18, Wang18,  Shen19, Fan19, Tang19, Matsuoka19}.
Intriguingly, these studies find that, in most cases, the Eddington ratios are close to the Eddington limit (i.e., $L_\mathrm{bol}/L_\mathrm{Edd}\sim1$), which indicates that the SMBHs at this epoch are in their most actively growing phase.
Such a high accretion rate seems to be a unique characteristic of the highest-redshift quasars because the observed Eddington ratio of $z\lesssim4$ quasars is $L_\mathrm{bol}/L_\mathrm{Edd}\sim0.01-0.1$ \citep[e.g.,][]{Shen08, Schulze15}.
On the other hand, some recent papers have identified less active (i.e., sub-Eddington) SMBHs at $z\gtrsim6$ \citep{Mazzucchelli17, KimY18, Shen19}.
\citet{Shen19} derive the Mg{\sc ii}-based black hole mass for 30 luminous quasars and argued that there is no significant difference in the Eddington ratio distribution of $z\gtrsim6$ quasars and lower-redshift quasars at the same luminosity range.
Therefore, it still is a matter of debate whether $z\gtrsim6$ quasars are an extremely active population.

Most of the previous studies have focused on the most luminous quasars at the reionization epoch, in other words, the most massive ($M_\mathrm{BH}\gtrsim10^9M_\odot$) and active ($L_\mathrm{bol}/L_\mathrm{Edd}\sim1$) black holes.
These studies do not include  
less massive SMBHs whose growth is in its most intense phase, and also SMBHs with smaller Eddington ratios, perhaps transitioning to a quiescent phase.
Therefore, mass measurements of $z\gtrsim6$ low-luminosity quasars have the potential to reveal a less biased view of early SMBH growth.
\citet[][hereafter W10]{Willott10a} presented the NIR properties of nine quasars found by the Canada-France High-$z$ Quasar Survey \citep[CFHQS;][]{Willott07}.
This is the only study to date to focus on the black hole masses of $z\gtrsim6$ low-luminosity quasars down to $L_\mathrm{bol}\sim10^{46.5}$ erg s$^{-1}$.

The SMBH properties of $z\gtrsim6$ low-luminosity quasars are also important in the context of galaxy-SMBH co-evolution.
\citet{Venemans16} showed that
three SMBHs at $z>6.5$ are ``over-massive" with respect to the dynamical mass of their host galaxies (as a proxy for stellar mass) when the mass ratio is compared to the bulge-to-BH mass ratio in the local universe \citep{KorHo13}.
\citet{Wang16} and \citet{Decarli18} find similar results for a sample of luminous quasars at $z\gtrsim6$.
However, it is possible that these objects are outliers in the build-up of the tight relation between galaxy bulge mass and central SMBH mass. 
Moreover, this overmassive trend is likely to be affected by luminosity bias; only the most massive SMBHs accreting at the Eddington limit will be luminous enough to enter the sample \citep[e.g.,][]{Schulze11}.

The Hyper Suprime-Cam (HSC) Subaru Strategic Program \citep[HSC-SSP;][]{HSCSSP} has covered $\sim 400$ deg$^2$ with five broad-band filters ($grizy$) to full depth ($5\sigma$ depth $z_\mathrm{lim, 5\sigma}=25.5$ mag) between 2014 Spring and 2019 March, powered by the wide field-of-view of the HSC (1.5 degrees in diameter; \citealt{Miyazaki18}).
With this survey, our team has recently discovered more than $80$ low-luminosity quasars at $z\sim6-7$ \citep{Matsuoka16, Matsuoka18a, Matsuoka18b, Matsuoka19}, with the highest-redshift one at $z=7.07$ \citep[HSC J1243+0100;][]{Matsuoka19}.
The number of the HSC quasars is already comparable to other major surveys such as the Sloan Digital Sky Survey (SDSS; \citealt{York00, Jiang16}) and the Panoramic Survey Telescope and Rapid Response System (Pan-STARRS1, PS1; \citealt{Kaiser10, Banados16}), but the luminosity range of the HSC quasars is  about an order of magnitude fainter than those luminous quasars, extending down to $M_{1450}\sim-22$ mag.
This unique sample of $z\sim6-7$ low-luminosity quasars now enables a study of less massive or less active SMBHs in the reionization epoch.

This is our sixth paper of the {\it Subaru High-$z$ Exploration of the Low-Luminosity Quasars (SHELLQs)} project and is the first to present the SMBH properties of our low-luminosity quasars.
In addition to the discovery papers, \citet[][Paper III]{Izumi18} and \citet[][Paper VIII]{Izumi19}  present ALMA follow-up of the four quasars, in which we argue that the star-formation activity of their hosts is not extreme as those of luminous quasars, and the SMBH-to-host mass ratio is comparable to that found in the local universe.
\citet[][Paper V]{Matsuoka18c} present the measurement of the quasar luminosity function at $z\sim6$.
In this paper, we present near-infrared spectroscopic observations of six HSC quasars with VLT/X-Shooter and Gemini/GNIRS, with a focus on virial black hole mass measurements.
The paper is organized as follows: 
The sample selection, observation, and data analysis are described in Section~\ref{sec:obs}.
The spectral fitting of the quasar continuum and the emission lines are described in Section~\ref{sec:spec_model}, followed by the mass measurements in Section~\ref{sec:BHmass}.
The implications for early SMBH growth from this work are discussed in Section~\ref{sec:MBHacc}.
Finally, Section~\ref{sec:summmary} gives the summary and our future prospects.
The magnitudes quoted in this paper are in the AB system.
We adopt a standard $\Lambda$CDM cosmology with $H_0=70$ km s$^{-1}$ Mpc$^{-1}$, $\Omega_m=0.3$, and $\Omega_\Lambda=0.7$.

\section{Data} \label{sec:obs}
We obtained near-infrared spectra of six HSC quasars with X-Shooter on VLT (UT2) and the Gemini Near-InfraRed Spectrograph (GNIRS) on Gemini-North.
Both instruments are medium-resolution echelle spectrographs, which enable us to observe various broad emission lines and the underlying continuum over the entire near-infrared wavelength range.
From the parent samples of low-luminosity quasars discovered in \citet{Matsuoka16} and \citet{Matsuoka18a}, we select targets for which Mg{\sc ii}-based mass measurements would be  feasible with ground-based 8m telescopes.
The optical redshift range is limited to $z_\mathrm{opt}\geq6.04$ to avoid severe atmospheric absorption at the Mg{\sc ii} line ($\lambda_\mathrm{obs}>1.97\micron$).
Also, the targets were limited to those whose absolute $1450$\AA\  magnitudes derived from their discovery spectra are $M_{1450}\leq-23.8$.
This magnitude cut is needed to keep the required integration times under ten hours.
Therefore, our objects are at the sensitivity limit of NIR follow-up observations.
We did not observe all objects satisfying these criteria, but gave highest priority to objects of higher redshift and luminosity.

Table~\ref{tab:BHmass_targets} summarizes our targets and observations.
Our observations were carried out in queue mode  between December 2016 and March 2018 at the VLT (Program ID: 098.A-0527) and between August 2016 and July 2017 at Gemini-North (Program ID: GN-2016B-FT-2, S17A0039N).
We observed three quasars at $6.37\leq z_\mathrm{opt}\lesssim 6.7$ with VLT/X-Shooter (J1205-0000, J0859+0022, J1152+0055)  and three quasars at $6.09\leq z_\mathrm{opt}\leq 6.26$ with Gemini/GNIRS (J2239+0207, J1208-0200, J2216-0016).
The absolute $1450$\AA\  magnitude of these quasars is roughly $2$ magnitudes on average fainter than \replaced{the most luminous $z>5.7$ quasars}{the luminous SDSS and PS1 quasars at $z>5.7$} \citep[e.g.,][]{Jiang16, Banados16}.
The six targets are also as faint as or even fainter than the moderately low-luminosity quasars presented in W10.
Five of the six HSC quasars in this paper (all but J1205-0000) were followed up with ALMA to measure their host properties ([C{\sc ii}] $158\micron$ emission line and dust continuum).
J0859+0022, J1152+0055, and J2216-0016 were observed in  ALMA Cycle 4  \citep{Izumi18}.
J2239+0207 and J1208-0200 were observed in ALMA Cycle 5, and will be presented in a forthcoming paper \citep{Izumi19}.

\begin{deluxetable*}{lCCClll}[htbp!]
\tablecaption{Our Sample and Observations \label{tab:BHmass_targets}}
\tablecolumns{7}
\tablenum{1}
\tablewidth{0pt}
\tablehead{
\colhead{ID} &
\colhead{$z_\mathrm{opt}$} &
\colhead{$y_{AB}$} &
\colhead{$M_\mathrm{1450, optical}$} &
\colhead{Date} &
\colhead{Instrument} &
\colhead{Exp. time} \\
\colhead{} &
\colhead{} &
\colhead{[mag]} &
\colhead{[mag]} &
\colhead{} &
\colhead{} &
\colhead{[hours]}
}
\startdata
HSC J120505.09--000027.9 & $6.7-6.9^a$ & $22.60\pm0.03$ & $-24.56\pm0.04$ & 2017 Mar. 31 &  VLT/X-Shooter & 7.2 (NIR)\\
&  & & &  \& 2018 Feb.  15, 16, 21&  & 6.5 (VIS)\\
HSC J085907.19+002255.9 & $6.39$ &  $23.63\pm0.07$ & $-24.09\pm0.09$ & 2016 Dec. 29& VLT/X-Shooter & 7.2 (NIR)\\
&  & & & \& Mar. 23,24,30,31 & & 6.5 (VIS) \\
&  & &  & \& 2017 Dec. 21 & & \\
HSC J115221.27+005536.6 & $6.37$ & $21.61\pm0.02$ & $-25.31\pm0.04$ & 2018 Mar. 12,13,14,15 & VLT/X-Shooter & 5.8 (NIR) \\
&  & & &    &  & 5.2 (VIS)\\
HSC J223947.47+020747.5 & $6.26$ &  $22.32\pm0.03$& $-24.69\pm0.04$ & 2016 Aug. 7, 8, 10,  & Gemini-N/GNIRS  &  2.7\\
 &  & &   & \& Sep. 27 &  &  \\
HSC J120859.23--020034.8& $6.2^a$ &  $22.05\pm0.03$& $-24.73\pm0.02$ & 2017 Feb. 17, 18, 19 &Gemini-N/GNIRS & 3.7\\
HSC J221644.47--001650.1& $6.09$ &  $22.96\pm0.04$& $-23.82\pm0.04$ & 2017 Jun. 25, 26, 28  &Gemini-N/GNIRS & 9.0\\
&   & & & \& Jul.  3, 5 & & \\
\enddata
\tablecomments{The optical redshift $z_\mathrm{opt}$ and the $1450$\AA\ magnitude quoted in this table are from our discovery papers \citep{Matsuoka16, Matsuoka18a}.
The $y$-band PSF magnitudes are from the latest HSC-SSP internal source catalog (DR S18A).
Galactic extinction is corrected \citep{Schlegel98}.
For the X-Shooter targets, we separately show the total exposure times in the NIR and VIS arms.
}
\tablenotetext{a}{The optical redshifts are uncertain due to an unusual spectrum (J1205-0000; \citealt{Matsuoka16} and Section~\ref{sec:J1205} in this paper) and due to an unsharp Lyman break (J1208-0200; \citealt{Matsuoka18a}). }
\end{deluxetable*}

\subsection{VLT/X-SHOOTER} \label{sec:obs_vlt}
\begin{figure*}[ht!]
\centering
 \includegraphics[width=\linewidth]{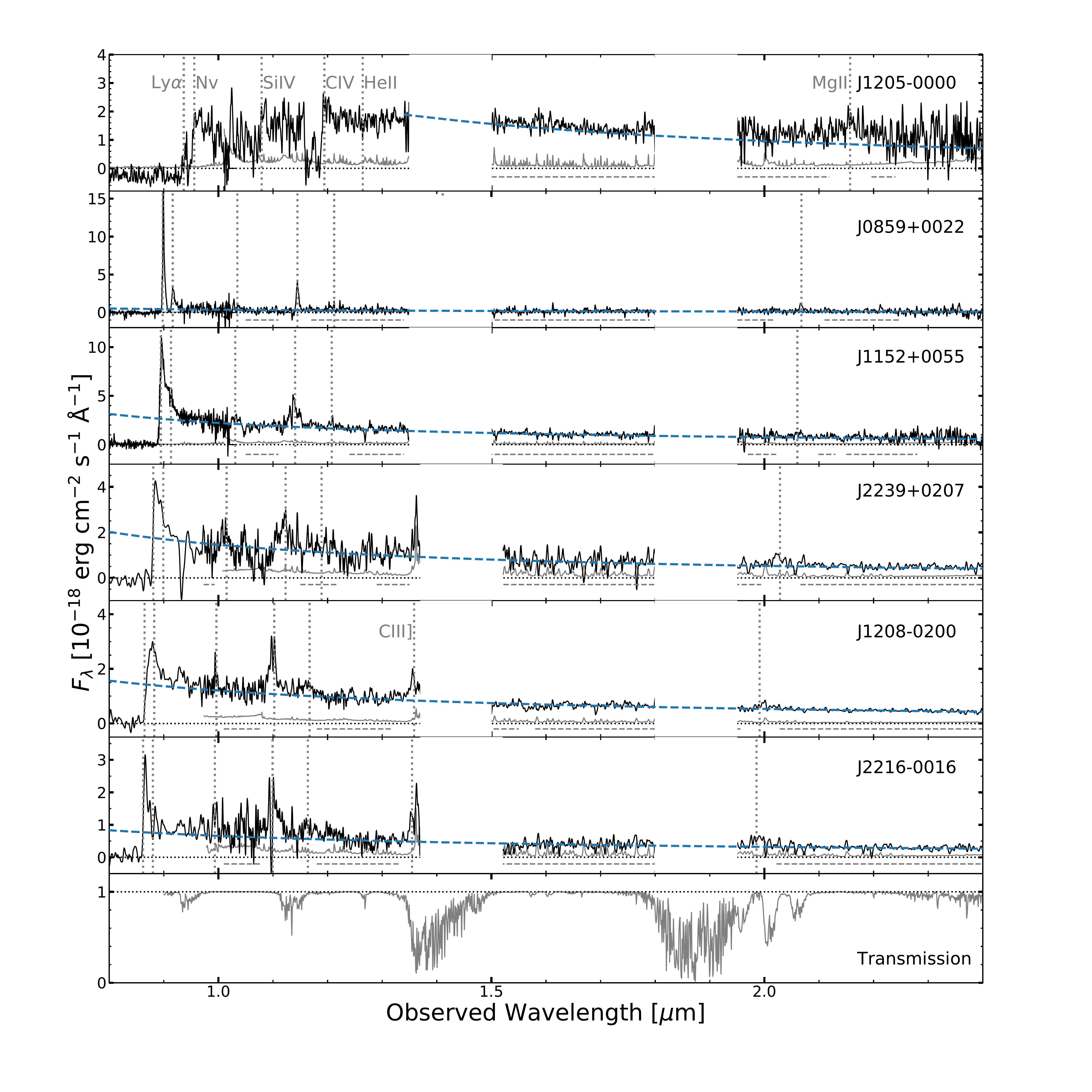}
\caption{Optical-to-NIR spectra of the six HSC quasars in this paper. 
The top three panels show the X-Shooter spectra of J1205-0000, J0859+0022, and J1152+0059.
The next three show the GNIRS spectra of J2239+0207, J1208-0200, and J2216-0016, while their optical data ($\lambda_\mathrm{obs}\lesssim1\micron$) are from the discovery spectra \citep{Matsuoka16, Matsuoka18a}. 
The bottom panel shows the atmospheric transmission at  Maunakea retrieved from the Gemini Observatory \citep{Lord92}.
The vertical lines indicate the central wavelength of broad emission lines expected from the systemic redshift (Mg{\sc ii} redshift for J1205-0000, and [C{\sc ii}] redshift from \citet{Izumi18, Izumi19} for the others).
In each panel, the noise spectra are shown in grey and the best-fit power-law continuum models are shown in blue.
} \label{fig:spec_vlt}
\end{figure*}

The VLT/X-Shooter consists of three echelle spectrographs 
covering different wavelength ranges: UVB ($\lambda_\mathrm{obs}=3000-5600$\AA), VIS ($\lambda_\mathrm{obs}=5500-10200$\AA), and NIR ($\lambda_\mathrm{obs}=10200-24800$\AA) arms.
We only observed with the VIS and NIR arms because $z\gtrsim6$ quasars have no signal in the UV arm due to strong absorption by the intergalactic medium.
The slit width was set to $0\arcsec.9$ in both arms.
The VIS arm was set to high-gain $2\times2$ pixel-binning  slow-readout mode.
This configuration results in a moderate spectral resolution of $R\sim 7410$ in the VIS arm and $R\sim 5410$ in the NIR arm.
Blind offsets from nearby bright stars were used for target acquisition.
The integration was divided into pairs of $\sim200$ second single exposures to subtract time-varying sky emission lines with the standard ABBA offset procedure.
The VLT targets were observed at an airmass of $\sim1.1-1.3$,  with moderate seeing ($\sim0.8$ arcsecond).
The data were processed with the dedicated ESO X-Shooter pipeline version 2.9.3 (Reflex).
The ABBA exposures were stacked.
Tracing the dispersion direction gave an averaged spatial profile, which we fit to a Gaussian to use as weights for the extraction.
A-type standard stars were observed for relative flux calibration.
Atmospheric absorption was corrected with sky transmission models obtained from {\sl SkyCalc} version 2.0.1 assuming the seasonal average at the observation dates and airmass\footnote{http://www.eso.org/sci/software/pipelines/skytools/}.
Absolute flux calibration for J0859+0022 and J1152+0055 was based on the HSC-$z$ band magnitudes in the internal S18A data release  ($z_{AB}=22.79\pm0.01$ and $z_{AB}=21.82\pm0.01$, respectively).
For J1205-0000, the $Ks$ magnitude of the VISTA Kilo-Degree Infrared Galaxy Survey \citep[VIKING][]{VIKING} was used.
After stacking the 1D spectra taken on different dates, we smoothed the extracted spectra with a gaussian kernel of $\sigma_\mathrm{kernel}=10$ pixels (FWHM $\sim 200$ km s$^{-1}$ at $K$-band) in the NIR arm and $\sigma_\mathrm{kernel}=15$ pixels (FWHM $\sim 200$ km s$^{-1}$ at $z$-band) in the VIS arm.
Figure~\ref{fig:spec_vlt} (top three panels) shows the optical-to-NIR spectra of the three VLT targets.

\subsection{Gemini/GNIRS} \label{sec:obs_gemini}
The GNIRS observations were carried out in cross-dispersed mode to cover the observed wavelengths of $\lambda_\mathrm{obs}\sim0.9-2.5\micron$, corresponding to the rest-frame wavelengths of $\lambda_\mathrm{rest}\approx0.12-0.35$ \micron.
We used the 31.7 l/mm grating and the short camera (0.15 arcsec per pixel).
The $0\arcsec.675$-slit was used for J1208-0200 and J2216-0016, which results in a spectral resolution of $R\sim760$.
A slightly wider $1\arcsec.0$-slit was used for J2239+0207 because the reference star for the blind offset acquisition was relatively far ($\sim40$ arcsec) from the quasar.
A spectral resolution of $R\sim510$ was achieved in this case.
The single exposure time was set to 300 sec with the standard ABBA nodding offset.
The observation was carried out in moderate weather conditions at airmass $\sim1.1-1.2$ and the seeing size of $\sim0.5-1.1$ arcsecond.
We observed nearby A-type stars at similar airmass every $\sim1-2$ hours during the observations to use their spectra to correct for atmospheric absorption.


The GNIRS data were reduced with the dedicated Gemini IRAF pipeline Version 1.13.
Following the standard procedure, we first flat-fielded the 2D spectra and subtracted sky background using each pair taken at the A and B nod positions.
The 2D spectra taken on each night were stacked to maximize the signal-to-noise ratio.
Distortion correction was done based on the GNIRS pinhole spectra.
After the 2D spectra were straightened in each order, the pixel-to-wavelength calibration was determined with the arc spectra.
The 1D extraction was performed in the same way as the X-Shooter spectrum.
The A-type star spectra were reduced in the same manner and used for relative flux calibration and atmospheric absorption correction.
The spectra observed on different dates were combined at this stage.
The absolute flux calibration was done by scaling the derived spectra to the broad-band photometry.
The HSC-$y$ magnitude was used for J2239+0207.
As the bluest order of the spectrum is too noisy for J2216-0016 and its average spatial profile cannot be determined,
we estimate the $J$-band magnitude by extrapolating the HSC-$y$ magnitude with a power-law continuum derived in the continuum fitting (Section~\ref{sec:spec_model}).
The GNIRS spectrum is scaled to match the expected $J$-band magnitude.
J1208-0200 is visible in VIKING DR3 images, but is not listed in the source catalog.
We measured its two-arcsecond aperture photometry in the $Ks$-band image ($Ks=21.66\pm0.13$) with {\it SExtractor} version $2.1.5$ \citep{SExtractor} and used this to scale the spectrum.
Finally, all the orders were combined into one spectrum and
the obtained spectra were smoothed with a Gaussian kernel with $\sigma_\mathrm{kernel}=1$ pixel (FWHM $\approx200$ km s$^{-1}$ at $K$-band).
Figure~\ref{fig:spec_vlt} (fourth to sixth panels) shows the near-infrared spectra of the three GNIRS targets.
In this figure, the optical data are from discovery spectra taken by Subaru/FOCAS \citep{Matsuoka16, Matsuoka18a}.


\section{Spectral Fitting} \label{sec:spec_model}
The derived near-infrared spectra are fitted with a multi-component continuum+emission line model.
We first subtract the power-law continuum and the iron emission line forest to measure the broad emission lines.
The Balmer continuum is also typically taken into account in similar analyses of the literature \citep{DeRosa14, Mazzucchelli17}, but we ignore this third component due to the relatively low spectrum quality.

\subsection{Continuum}  \label{sec:spec_model_cont}
The most dominant component of the rest-frame UV spectrum of an unobscured quasar is the power-law continuum.
Here, we model the observed continuum with a single power-law: 
\begin{equation}
 F_\mathrm{\lambda, cont}=F_0 \ \lambda^{\alpha_\lambda},
\end{equation}
where $F_0$ is the scaling factor and $\alpha_\lambda$ is the power-law index.
The slope of the quasar continuum is known not to evolve with redshift, with the most commonly used values in the literature of  $\alpha_\lambda=-1.5$.
\citet{Selsing16} found a slope of $\alpha_\lambda=-1.70$ from their composite X-Shooter spectrum of $1<z<2$ quasars.
At higher redshift, \citet{Cristiani16} compiled 1669 luminous quasars at $3.6<z<4.0$, showing a median slope of $\alpha_\lambda=-1.36$ with a dispersion of $0.36$.

The secondary component is the forest of iron emission lines.
These lines are observed as a pseudo-continuum because iron has thousands of blended weak emission lines at $\lambda_\mathrm{rest}\approx2000-3000$\AA, which affects the measurement of the Mg{\sc ii} line.
In this work, we use the empirical template of \citet{Vestergaard01}, which is derived from a high-resolution spectrum of I Zw I, a nearby Seyfert 1 galaxy at $z=0.061$ with relatively narrow and strong iron emission lines.
While there is no flux at the Mg{\sc ii} region in the \citet{Vestergaard01} iron template, we added constant flux in their template at $\lambda_\mathrm{rest}=2770-2820$\AA,  to include unidentified iron emission underneath Mg{\sc ii}.
We follow \citet{Kurk07} and add $20$\% of the mean continuum flux density at $\lambda_\mathrm{rest}=2930-2970$\AA.
The narrow iron emission lines are smoothed with a Gaussian kernel in logarithmic wavelength space with
\begin{equation}
\sigma_\mathrm{conv}=\sqrt{\mathrm{FWHM}^2_\mathrm{QSO} - \mathrm{FWHM}^2_\mathrm{I\ Zw\ I} } / 2\sqrt{2 \ln{2}},
\end{equation}
where $\mathrm{FWHM}_\mathrm{I\ Zw\ I}=900$ km s$^{-1}$ is the line width of the I Zw I template and $\mathrm{FWHM}_\mathrm{QSO}$ is the line width of the target quasar.
Three broadened iron templates with $\mathrm{FWHM}_\mathrm{QSO}=1000$, $2500$, and $5000$ km s$^{-1}$ are generated and used in the spectral fitting.

We estimate the relative contribution of the power-law and iron components with an iterative process \citep{Vestergaard01}.
First, a power-law is fitted to selected regions in the spectra.
We avoid the order gaps and regions where emission and absorption lines are visible; thus the fitted regions vary from target to target.
In the initial fitting, only the cleanest range is selected.
Second, after subtracting the estimated power-law model, the iron templates are fitted to the residuals by scaling the templates.
The fitting windows for the iron templates are $\lambda_\mathrm{rest}=2200-2750$\AA\ and $\lambda_\mathrm{rest}=2850-3090$\AA, while we manually removed regions where atmospheric absorption is severe and where the continuum is not smooth\footnote{We see non-smooth continuum at $\lambda_\mathrm{obs}\gtrsim2.3\micron$ in the VLT/X-Shooter spectra. This region may be affected by thermal noise in the instrument.}.
The best iron templates with the smallest residuals are chosen at this stage.
After determining the scale of the iron templates, the best-fit iron models are subtracted from the original spectra,
and the power-law continuum is fitted again with a larger fitting window.
Most of the continuum range is included in this step.
Then, the iron template is fitted again after subtracting the newly determined power-law continuum model.
This iterative fitting is repeated until achieving convergence of $<1\%$.
The result of this continuum measurement for each quasar is summarized in Table~\ref{tab:shellqs_cont}.
The range of the power-law index is $-1.58\leq\alpha_\lambda\leq-1.04$ except J1205-0000 (Section~\ref{sec:J1205}).
Since this range is consistent with typical values for type-I quasars,
we assume that the host galaxy contribution to the observed continuum is negligible.
The best-fit continuum (power-law + iron) models are shown in Figure~\ref{fig:spec_vlt}.
It is noted that only the near-infrared spectra ($\lambda_\mathrm{obs}\gtrsim1\micron$) are used to determine the continuum models, but 
the best-fit power-law models fit the optical spectra well.

The uncertainties of the continuum parameters are measured by a Monte Carlo approach \citep{Shen11}.
From each unsmoothed flux-calibrated spectrum, $100$ mock spectra are generated by adding random noise to each spectral pixel based on its noise vectors.
The mock spectrum is then smoothed by the same Gaussian kernels and the power-law+iron pseudo-continuum are fitted with the same procedure as the original spectrum.
The $1\sigma$ uncertainty is given by $16$\% and $84$\% percentiles of the distribution of the best-fit values.
Finally, we update the rest-frame $1450$\AA\ magnitude with the best-fit power-law continuum models and their uncertainties, as reported in Table~\ref{tab:shellqs_cont}.
Note that the absolute continuum magnitude of J0859+0022 is about one magnitude fainter than the value quoted by \citet{Matsuoka18a}. 
This discrepancy is at least partly due to the low quality of the spectrum used in that paper (a $30$ minutes exposure with Subaru/FOCAS) and their different and narrower fitting window ($\lambda_\mathrm{rest}=1265-1345$ \AA) than that of this paper ($\lambda_\mathrm{rest}=1420-3040$ \AA). 

\begin{deluxetable*}{LCCCCCC}[htbp!]
\tablecaption{Best-fit Continuum Parameters \label{tab:shellqs_cont}}
\tablecolumns{7}
\tablenum{2}
\tablewidth{0pt}
\tablehead{
\colhead{} &
\colhead{J1205-0000$^a$} &
\colhead{J0859+0022} &
\colhead{J1152+0055} &
\colhead{J2239+0207} &
\colhead{J1208-0200} &
\colhead{J2216-0016} 
}
\startdata
\alpha_\lambda  & -1.69_{-0.21}^{+0.14} & -1.58_{-0.22}^{+0.14} & -1.49_{-0.03}^{+0.02} & -1.47_{-0.05}^{+0.06} & -1.18_{-0.02}^{+0.02} & -1.04_{-0.06}^{+0.08}   \\
 $M_{1450}$ & $-25.54\pm 0.28$& $-23.10\pm 0.27$ & $-25.08\pm 0.07$ & $-24.60\pm0.15$ & $-24.36\pm0.09$ & $-23.65\pm0.20$   \\
\enddata
\tablenotetext{a}{For J1205-0000, we use only the $H$- and $K$-band spectra for the continuum fitting. See Figure~\ref{fig:spec_vlt} and Section~\ref{sec:J1205}.}
\end{deluxetable*}

\subsection{Emission Lines and Redshifts} \label{sec:spec_model_line}
\begin{figure*}[ht!]
\plotone{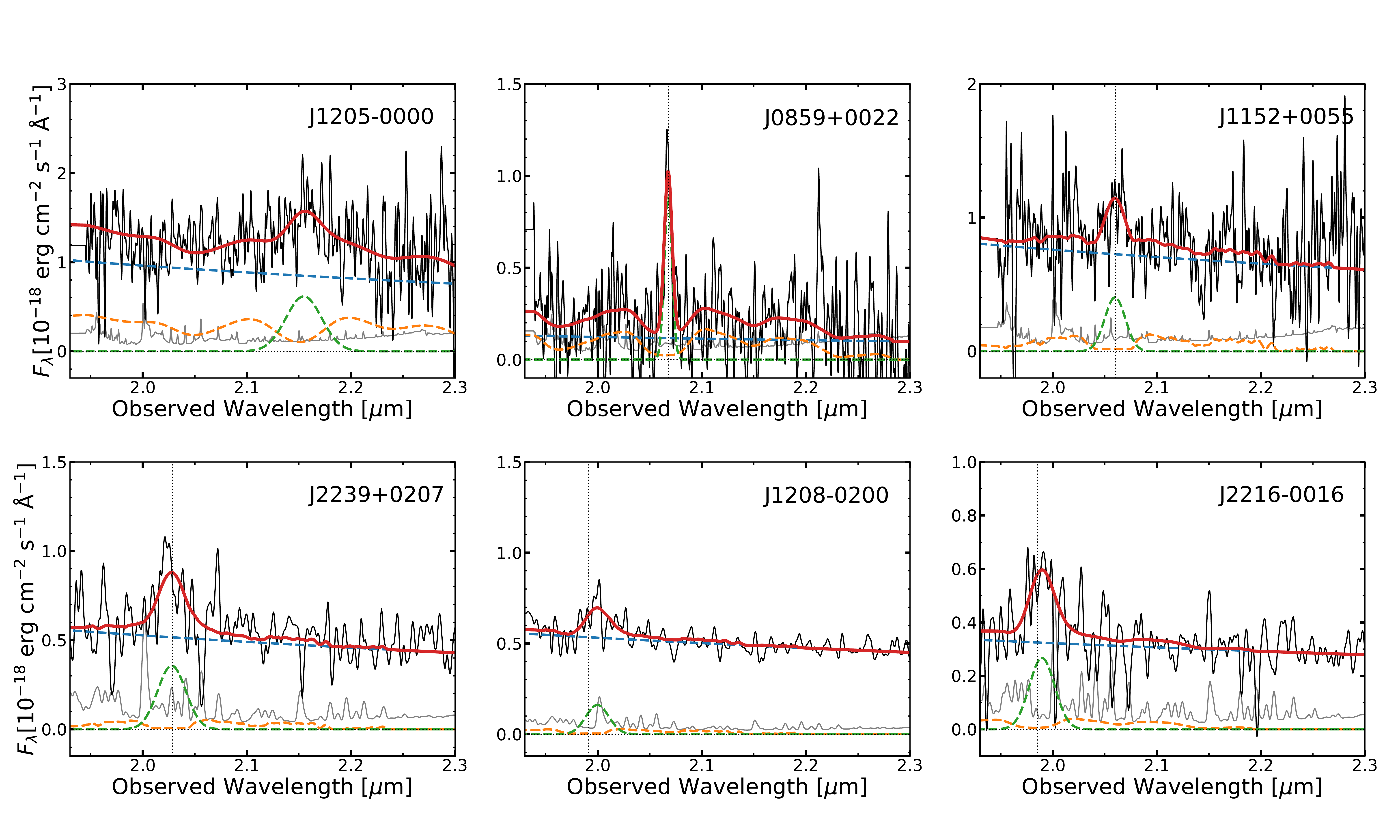}
\caption{The best-fit continuum+line models around Mg{\sc ii}.
The best-fit models are shown in blue (power law), orange (iron), green (Gaussian), and red (sum of the three models).
The vertical lines show the systemic redshift obtained from [C{\sc ii}] emission lines \citep{Izumi18, Izumi19}.
}.\label{fig:MgII}
\end{figure*}

\begin{figure*}[ht!]
\plotone{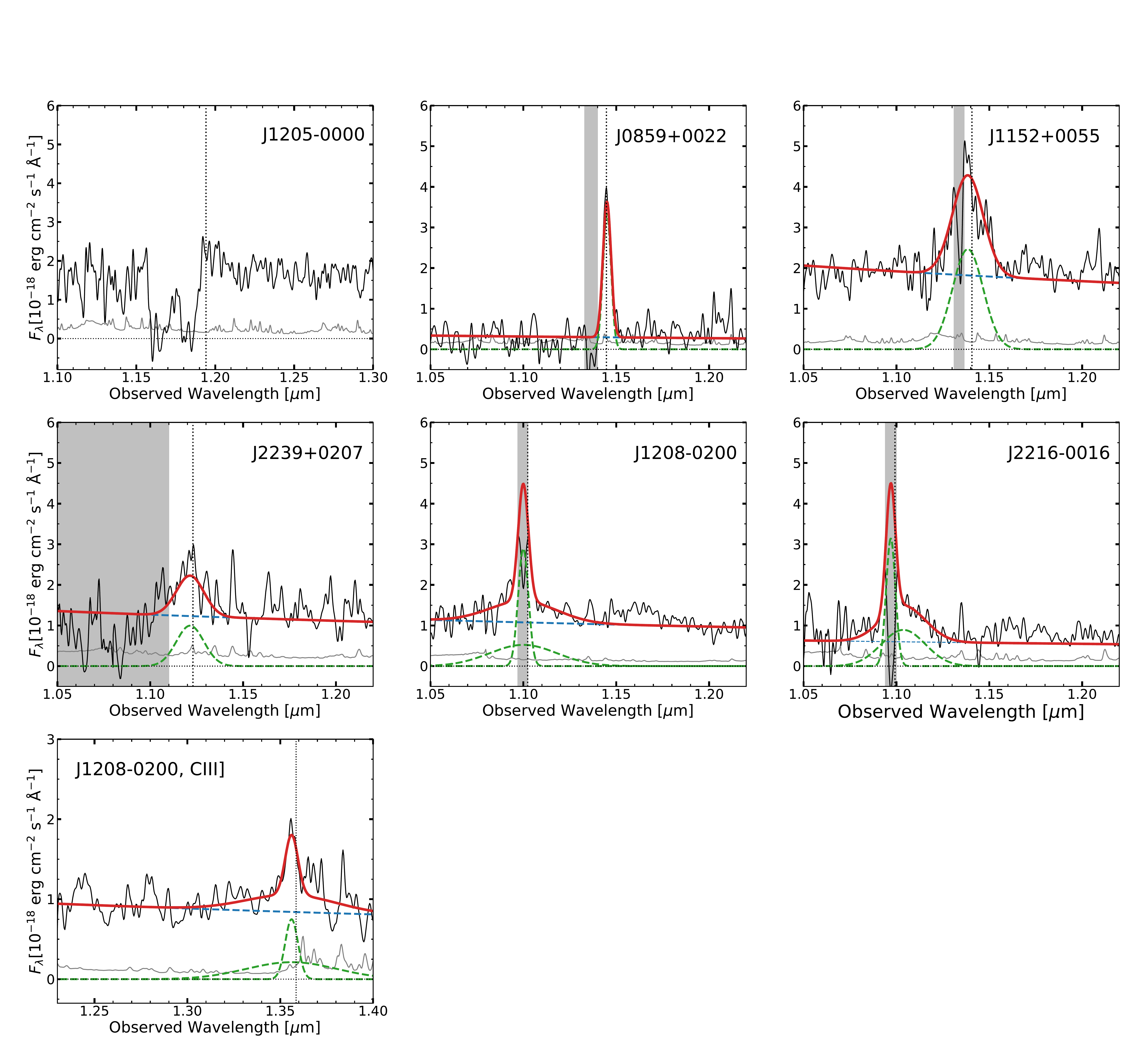}
\caption{The best-fit continuum+line model around C{\sc iv} and C{\sc iii]} (J1208-0200).
Shading shows the masked regions in the emission line fitting.
}.\label{fig:CIV}
\end{figure*}

We fit the strong UV emission lines (C{\sc iv} $\lambda 1549$, C{\sc iii]} $\lambda 1909$, and Mg{\sc ii} $\lambda 2798$) after subtracting the best-fit continuum (power-law and iron emission lines) models.
We detect broad Mg{\sc ii} and C{\sc iv} emission lines for all six quasars, while C{\sc iii]} is also detected in J1208-0200 (and tentatively in J2239+0207 and J2216-0016).
There are also other weaker lines visible in the near-infrared spectra: S{\sc iv} $\lambda 1397$ (J1205-0000, J1152+0055, J2239+0207, J1208-0200) and He{\sc ii} $\lambda 1640$ (J1152+0055, J1208-0200, J2216-0016).
The strong emission lines are fitted with a Gaussian profile with the free parameters of scaling factor, central wavelength, and line width.
We use a single Gaussian in most cases, but fit a second Gaussian when the emission lines have broad line skirts or asymmetric profiles which cannot be well fitted with a single Gaussian.
The central wavelengths of the two Gaussian components are fixed at the same positions in all cases except the C{\sc iv} emission line of J2216-0016, which shows significant asymmetry.
We measure the Mg{\sc ii} redshift from the best-fit Gaussian profile.
For J1205-0000, we shift the iron templates assuming the Mg{\sc ii} redshift and repeat the continuum+line fitting until the best-fit Gaussian returns the same redshift.
We also measure blueshifts with respect to [C{\sc ii}] redshift for Mg{\sc ii}, C{\sc iv}, and C{\sc iii]}.
The uncertainties of the line profiles are measured with the $100$ mock spectra that we use in measuring the uncertainties of continuum parameters.
For each mock spectrum, we subtract the best-fit continuum model and fit single or double Gaussian profiles to the residuals,
in order to take into account the effects that the continuum estimate potentially has on the line profile measurements.
The derived emission line properties including the redshift measurements are shown in Table~\ref{tab:shellqs_lineprofiles}.
Figure~\ref{fig:MgII} and Figure~\ref{fig:CIV} show our best-fit continuum+line models around Mg{\sc ii} and C{\sc iv}, respectively.
The C{\sc iii]} line of J1208-02000 is also shown in Figure~\ref{fig:CIV}.

Intriguingly, we found C{\sc iv} absorption lines either at the line peak (J0859+0022, J1152+0055, J1208-0200, J2216-0016) or in the blue wing (J1205-0000, J2239+0207) in all six of our sources.
These absorption regions are masked in the line measurements (Figure~\ref{fig:CIV}).
This high BAL or associated absorption fraction seems to be higher than that both of more luminous $z>5.8$ quasars and also lower redshift quasars \citep[$\lesssim16\%$ for BALs;][]{Shen19, Maiolino04}.
Our sample is small, however; observations of additional low-luminosity $z\sim6$ quasars are needed to see if this trend continues.


We adopt the [C{\sc ii}] $158\micron$ redshifts as the systemic redshifts, which were measured in \citet{Izumi18} for J0859+0022, J1152+0055, and J2216-0016, and \citet{Izumi19} for J2239+0207 and J1208-0200.
We use the Mg{\sc ii} redshift for J1205-0000 as a proxy for its systemic redshift, because this quasar has not yet been observed with ALMA.
It is known that the C{\sc iv} $\lambda1549$ emission line is frequently blueshifted \citep[e.g.,][]{Shen16, Coatman17}.
Among the HSC quasars in this study, we find a significant C{\sc iv} blueshift ($\approx400-600\ \mathrm{km\ s^{-1}}$) for four quasars (J1152+0055, J2239+0207, J1208-0200, J2216-0016), while J0859+0022 shows a slightly {\it redshifted} C{\sc iv}.
Given that the flux density in the blue wing of C{\sc iv}  is around zero, which we mask in the line fitting, the C{\sc iv} emission line of J0859+0022 maybe affected by the BAL absorption, which would shift the observed peak of the emission line to longer wavelengths, giving a negative blueshift.
The C{\sc iv} velocity offsets from Mg{\sc ii} are $<2000$ km s$^{-1}$, which is broadly consistent with what is found for luminous $z\sim6$ quasars \citep{Meyer19, Shen19}\footnote{\citet{Mazzucchelli17} show larger Mg{\sc ii}-C{\sc iv} blueshifts for $z\gtrsim6.5$ quasars, as pointed out by \citet[][Sec.~4]{Shen19}.}.
The C{\sc iv} emission line of J2216-0016 has broad and narrow components with  different central wavelengths, while its narrow peak is affected by an absorption line.
We determine its C{\sc iv} blueshift $\Delta v_\mathrm{CIV-[CII], narrow}=630_{-120}^{+90}$ from its best-fit narrow Gaussian (FWHM$=1520^{+340}_{-190}$ km s$^{-1}$), while the central wavelength of its broad component is somewhat redder, corresponding to $\Delta v_\mathrm{CIV-[CII], broad}=-1220_{-760}^{+500}$ km s$^{-1}$ (FWHM$=7540^{+1880}_{-1460}$ km s$^{-1}$).

\citet{Wang16} find that, at $z\sim6$, even Mg{\sc ii} is often blueshifted ($\sim1000\ \mathrm{km s^{-1}}$) with respect to [C{\sc ii}] redshift.
However, we do not find such large Mg{\sc ii} blueshifts for any of the HSC quasars ($\lesssim200$ km s$^{-1}$).
The Mg{\sc ii} emission line of J1208-0200 is {\it redshifted} relative to [C{\sc ii}] by $\Delta v_\mathrm{Mg{\sc II}-[C{\sc II}]}=-1260^{+430}_{-350}$ km s$^{-1}$.
This Mg{\sc ii} offset may be affected by OH sky emission line residuals around the line peak ($\lambda_\mathrm{obs} \sim2.00\ \micron$; Figure~\ref{fig:CIV}).

\begin{deluxetable*}{lCCCCCC}[htbp!]
\tablecaption{Emission Line Parameters and Redshifts \label{tab:shellqs_lineprofiles}}
\tablecolumns{7}
\tablenum{3}
\tablewidth{0pt}
\tablehead{
\colhead{} &
\colhead{J1205-0000} &
\colhead{J0859+0022} &
\colhead{J1152+0055} &
\colhead{J2239+0207} &
\colhead{J1208-0200} &
\colhead{J2216-0016} 
}
\startdata
FWHM$_\mathrm{MgII}$ (km s$^{-1}$) & $5620_{-990}^{+220}$ & $1280_{-410}^{+240}$& $3240_{-450}^{+280}$ & $4670_{-700}^{+910}$ & $3850_{-1990}^{+920}$ & $4320_{-1060}^{+830}$ \\
EW$_\mathrm{MgII}$ (\AA)  & $41_{-6}^{+1}$ & $101_{-14}^{+12}$ & $17_{-2}^{+2}$ & $32_{-4}^{+4}$ & $12_{-4}^{+3}$ & $36_{-6}^{+7}$ \\
FWHM$_\mathrm{CIII], broad}$ (km s$^{-1}$) & $\cdots$  & $\cdots$ & $\cdots$& $\cdots$ & $12700_{-1800}^{+2700}$& $\cdots$ \\
FWHM$_\mathrm{CIII], narrow}$ (km s$^{-1}$) & $\cdots$  & $\cdots$ & $\cdots$& $\cdots$ & $1720_{-750}^{+880}$  & $\cdots$ \\
EW$_\mathrm{CIII], broad}$ (\AA)  & $\cdots$ & $\cdots$ & $\cdots$& $\cdots$ & $22_{-4}^{+5}$ & $\cdots$ \\
EW$_\mathrm{CIII], narrow}$ (\AA)  & $\cdots$ & $\cdots$ & $\cdots$& $\cdots$ & $11_{-3}^{+4}$ & $\cdots$ \\
FWHM$_\mathrm{CIV, broad}$ (km s$^{-1}$) & $\cdots$ & $1350_{-50}^{+110, \dagger}$ & $5190_{-360}^{+290, \dagger}$ & $4630_{-1260}^{+1040, \dagger}$ & $11600_{-2200}^{+3600, \dagger}$ & $7540_{-1460}^{+1880, \dagger}$ \\
FWHM$_\mathrm{CIV, narrow}$ (km s$^{-1}$) & $\cdots$ & $\cdots$ & $\cdots$ & $\cdots$ & $1680_{-360}^{+480, \dagger}$ & $1520_{-190}^{+340, \dagger}$ \\
EW$_\mathrm{CIV, broad}$ (\AA) & $\cdots$ & $87_{-6}^{+4, \dagger}$  & $39_{-2}^{+2, \dagger}$ & $21_{-3}^{+3, \dagger}$ & $31_{-4}^{+5, \dagger}$ & $62_{-12}^{+7, \dagger}$ \\
EW$_\mathrm{CIV, narrow}$ (\AA) & $\cdots$ & $\cdots$  & $\cdots$ & $\cdots$ & $26_{-2}^{+5, \dagger}$ & $47_{-7}^{+13, \dagger}$ \\
$\Delta v_\mathrm{Mg{\sc II}-[CII]}$ (km s$^{-1}$) & $\cdots$ & $110_{-70}^{+70}$  & $140_{-170}^{+160}$& $200_{-320}^{+270}$ & $-1260_{-350}^{+430}$ & $-540_{-300}^{+330}$ \\
$\Delta v_\mathrm{C{\sc III]}-[CII]}$ (km s$^{-1}$) & $\cdots$ & $\cdots$  & $\cdots$& $\cdots$ & $410_{-140}^{+140}$ & $\cdots$ \\
$\Delta v_\mathrm{C{\sc IV}-[CII]}$ (km s$^{-1}$) & $\cdots$ & $-60_{-41}^{+17}$  & $590_{-190}^{+130}$& $430_{-350}^{+410}$ & $570_{-60}^{+90}$ & $630_{-120}^{+90}$ \\
{$z_\mathrm{Mg{\sc II}}$} & $6.699_{-0.001}^{+0.007,*}$ & $6.388_{-0.002}^{+0.002}$  & $6.360_{-0.004}^{+0.004}$& $6.245_{-0.007}^{+0.008}$ & $6.144_{-0.010}^{+0.008}$ & $6.109_{-0.008}^{+0.007}$ \\
{$z_\mathrm{[CII]}$} & $\cdots$ & $6.3903^*$  & $6.3637^*$  & $6.2499^*$ & $6.1165^*$ & $6.0962^*$ \\
\enddata 
\tablecomments{
The line widths are corrected for instrumental broadening.
The equivalent widths are given in the rest frame values.
For the velocity offsets of emission lines, blueshifts are shown in positive values.
}
\tablenotetext{*}{We use the [C{\sc ii}] redshift as the systemic redshift for five objects, while we use the Mg{\sc ii} redshift for J1205-0000.}
\tablenotetext{\dagger}{Emission lines affected by absorption lines.}
\end{deluxetable*}

\subsection{Notes on Individual Objects}\label{sec:J1205_J0859}
Here we look at two quasars which show peculiar spectral properties: J1205-0000 (Section~\ref{sec:J1205}) and J0859+0022 (Section~\ref{sec:J0859}).

\subsubsection{J1205-0000}\label{sec:J1205}
J1205-0000 shows a remarkably flat continuum in its X-Shooter spectrum.
This quasar is detected in near-infrared surveys despite its faintness in the optical ($y=22.61$), as was pointed out by \citet{Mazzucchelli17}.
The two-arcsecond aperture magnitudes in the VIKING DR3\footnote{ESO Catalogue Facility: https://www.eso.org/qi/} are \replaced{$J_\mathrm{AB}=21.92\pm0.21$, $H_\mathrm{AB}=21.46\pm0.34$, and $Ks_\mathrm{AB}=20.72\pm0.18$}{$J=21.92\pm0.21$, $H=21.46\pm0.34$, and $Ks=20.72\pm0.18$}.
J1205-0000 is even detected in ALLWISE\footnote{data release November 13, 2013} with \replaced{$W1_\mathrm{AB}=19.98\pm0.15$ and  $W2_\mathrm{AB}=19.65\pm0.23$}{$W1=19.98\pm0.15$ and  $W2=19.65\pm0.23$}.
Thus, $y-W2$ is $2.96\pm0.23$.


Those spectral and broad-band properties of J1205-0000 can be attributed to its obscured nature.
\citet{Mazzucchelli17} estimate the dust reddening of this object as $E(B-V)=0.3$ (or $A_V=0.9$ mag, assuming the Galactic mean value of $R_V=3.1$ \citep{Cardelli89}.
They fit the composite spectrum of low-redshift quasars derived in \citet{Selsing16} to the broad-band SED ($J,H,Ks,W1,W2$) with the Calzetti extinction law \citep{Calzetti00}.
However, we noticed that J1205-0000 has two nearby sources visible in the HSC-$y$ image.
They are close enough that they likely contaminate the WISE flux of J1205-0000.
Such contamination would cause to overestimate the amount of dust extinction.
Our WISE flux decomposition and estimate of dust reddening of J1205-0000 will be fully described in a forthcoming paper \citep{Kato19}.

%

To measure the Mg{\sc ii} emission line of J1205-0000 in this paper, 
we restrict the power-law continuum fitting window to $\lambda_\mathrm{obs}>1.5\micron$.
For the iron emission lines, we use the same fitting window as the other HSC quasars.
We extrapolate the best-fit power-law continuum ($\alpha_\lambda=-1.69^{+0.14}_{-0.21}$) to determine an absolute $1450$\AA\ magnitude $M_{1450}=-25.54\pm0.28$.
The Mg{\sc ii} line profile is measured after subtracting the best-fit power-law continuum plus iron emission line models determined in this way.
Our best-fit Mg{\sc ii} profile ($z_\mathrm{MgII}=6.699^{+0.007}_{-0.001}$, FWHM$=5620^{+220}_{-990}$ km s$^{-1}$) is significantly different from that of  \citet[][$z_\mathrm{MgII}=6.73$, FWHM$=8841$ km s$^{-1}$]{Mazzucchelli17} based on a $4$ hour integration with the 6.5m-Magellan FIRE instrument.
This difference is likely due to confusion of the true Mg{\sc ii} emission line with the underlying iron emission in their noisy spectrum.

\begin{figure}[tb]
 \plotone{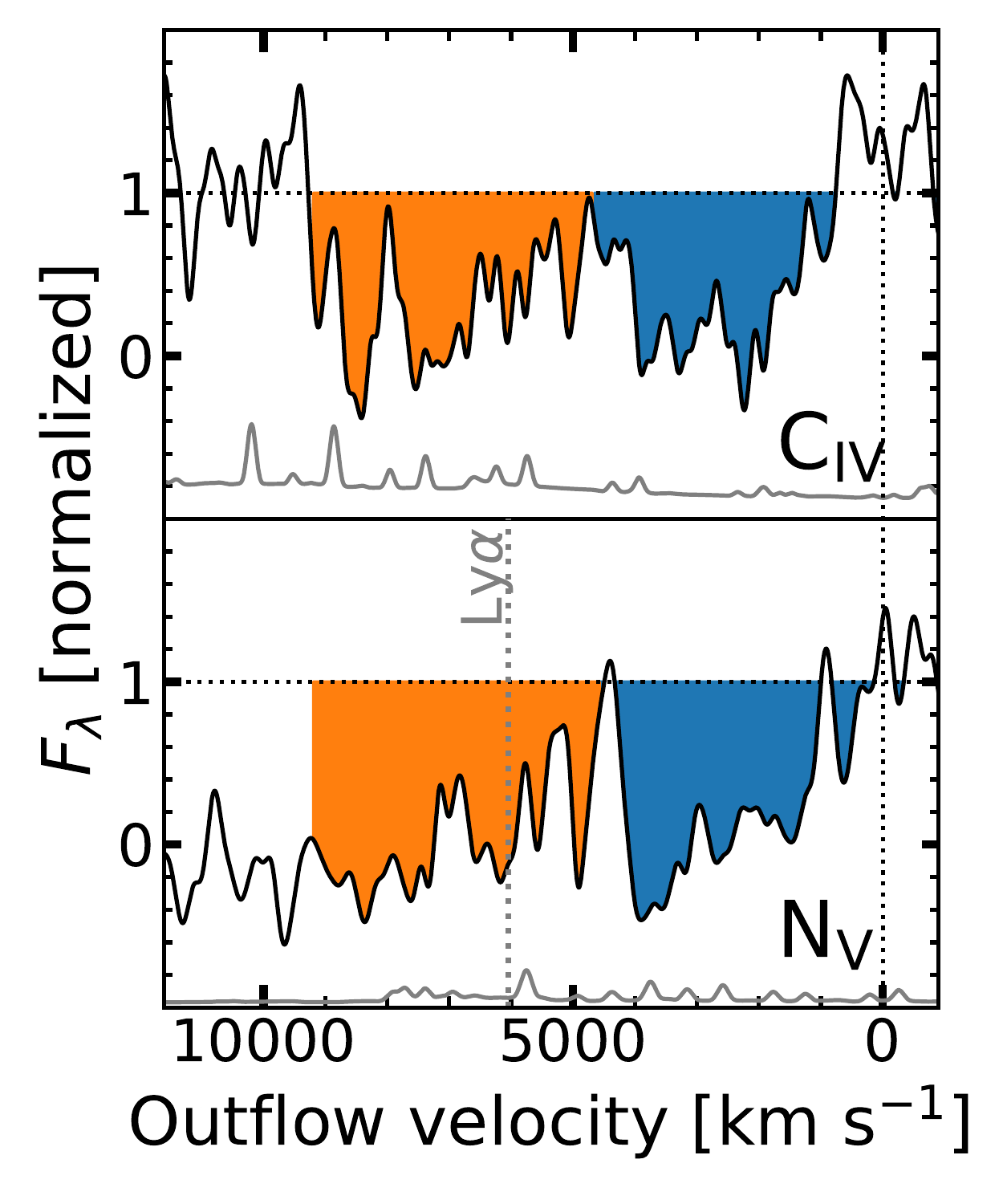}
 \caption{Broad absorption lines of J1205-0000.
 The top and bottom panels show the normalized spectrum around C{\sc iv} from the NIR arm and N{\sc v} from the VIS arm, respectively, with the wavelength converted to outflow velocity. 
The NIR arm spectrum is smoothed with a Gaussian kernel with $\sigma_\mathrm{kernel}=5$ pixels (FWHM$\sim180$ km s$^{-1}$ at C{\sc iv}) , while the VIS arm spectrum is smoothed with $\sigma_\mathrm{kernel}=15$ pixels (FWHM$\sim220$ km s$^{-1}$ at N{\sc v}). 
The two outflow components are shown in blue and orange for each line.
The vertical line at $6050$ km s$^{-1}$ in the bottom panel corresponds to the central wavelength of Ly$\alpha$ expected from the Mg{\sc ii} redshift.
 }.\label{fig:J1205_abs}
 \end{figure}

\citet{Matsuoka16} were unable to determine a reliable optical redshift of J1205-0000 from its Subaru/FOCAS spectrum ($z_\mathrm{opt}=6.7-6.9$) because of its unusual spectral shape around the Lyman break.
The Mg{\sc ii} detection in the deep X-Shooter spectrum gives a more precise measurement of the systemic redshift, $z_\mathrm{MgII}=6.699_{-0.001}^{+0.007}$. 
As seen in Figure~\ref{fig:spec_vlt}, this quasar has strong broad absorption lines (BALs) of high ionization lines (C{\sc iv}, Si{\sc iv}, and N{\sc v}).
Dust-reddened quasars seem to have a higher BAL fraction than do unobscured quasars \citep[e.g.,][]{Richards03}.
The BAL features of J1205-0000 are also visible in the Magellan/FIRE spectrum of \citet{Mazzucchelli17}.
J1205-0000 is one of the highest-redshift BAL quasars known to date, behind DELS J0038-1527 at $z=7.02$ \citep{Wang18} and also possibly HSC J1243+0100 at $z=7.07$ \citep{Matsuoka19}.
Figure~\ref{fig:J1205_abs} shows the outflow velocity of C{\sc iv} and N{\sc v}, in which the spectrum is normalized by constant flux.
The normalization scale is determined redward of each emission line.
There are two strong outflow components of C{\sc iv} BALs, with the first (system \#1) having a velocity of $\sim2900$ km s$^{-1}$ and the second  (system \#2) having $\sim7400$ km s$^{-1}$.
While not robust, those absorption lines may be further split into two components ($\sim1000$ and $\sim4000$  km s$^{-1}$ for system \#1 and $\sim7000$ and $\sim8000$  km s$^{-1}$ for system \#2) .
These (at least) two systems absorb Ly$\alpha$ emission, making it invisible (Figure~\ref{fig:J1205_abs}).
Extreme BAL quasars like J1205-0000 have been found in lower-redshift dust-reddened quasars \citep{Maiolino04, Hall02, Ross15}.


\subsubsection{J0859+0022}\label{sec:J0859}
J0859+0022 is the faintest of the six HSC quasars, with a continuum absolute magnitude ($M_{1450}=-23.10\pm0.27$) at the intersection of the luminosity functions of quasars and Lyman break galaxies at $z\sim6$ \citep{Onoue17, Matsuoka18c}.
Its near-infrared X-Shooter spectrum shows a weak continuum with strong and narrow emission lines (FWHM$_\mathrm{MgII}=1280_{-410}^{+240}$ km s$^{-1}$, FWHM$_\mathrm{CIV}=1350_{-50}^{+110}$ km s$^{-1}$).
In the optical spectrum of the X-Shooter, the red part of the Ly$\alpha$ emission line is well fitted by a double Gaussian with $\mathrm{FWHM}_\mathrm{Ly\alpha, broad}=2208_{-3}^{+14}$ km s$^{-1}$ and $\mathrm{FWHM}_\mathrm{Ly\alpha, narrow}=336_{-1}^{+1}$ km s$^{-1}$.
The N{\sc v} emission line is fitted by a single Gaussian with $\mathrm{FWHM}_\mathrm{Nv}=1787_{-4}^{+25}$ km s$^{-1}$.
We measure the Ly$\alpha$ + N{\sc v} rest-frame equivalent width EW$_\mathrm{Ly\alpha + N{\sc v}}=252$\AA\  ($\log{\mathrm{EW} }=2.40$\AA) by integrating the flux above the best-fit power-law continuum at $\lambda_\mathrm{rest}=1160-1290$\AA\ \citep{Diamond-Stanic09}.
The EW distribution is log normal, with a mean of $\mathrm{\left<\log{EW_{Ly\alpha + N{\sc v}}(\AA)}\right>}=1.542$ and dispersion of $\sigma (\log \mathrm{EW_{Ly\alpha + N{\sc v}} (\AA)})=0.391$  \citep{Banados16}.
Therefore, the Ly$\alpha$ + N{\sc v} rest-frame EW of J0859+0022 deviates from the average by $2.2\sigma$.

\begin{figure*}[t!]
 \plotone{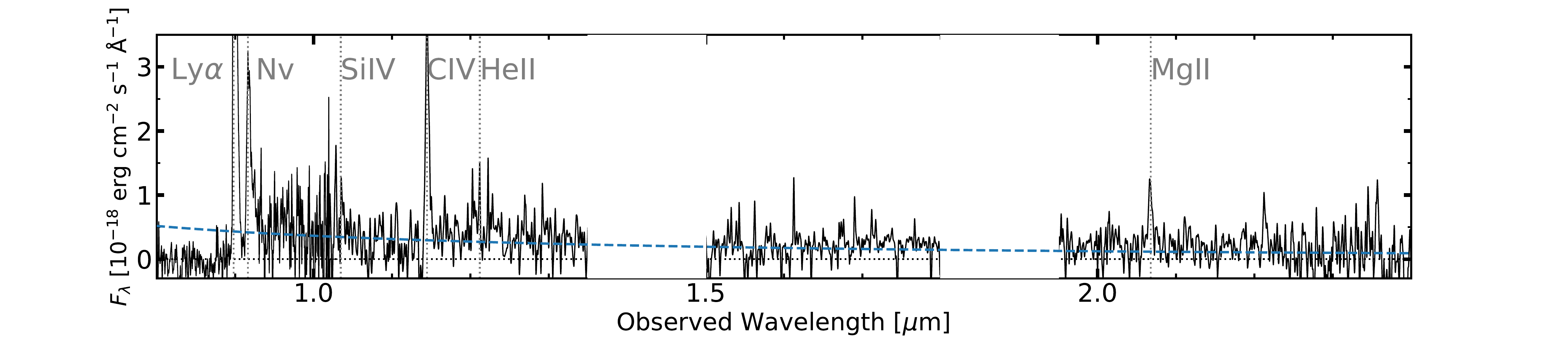}
 \caption{The continuum of J0859+0022.
 The NIR spectrum is smoothed with a FWHM$=20$ pixel Gaussian kernel.}.\label{fig:J0859_zoom}
 \end{figure*}

Those emission line properties are reminiscent of type-II quasars; however,
we assume that J0859+0022 is a low-luminosity type-I AGN for the following reasons.
First, the blue continuum slope of J0859+0022 ($\alpha_\lambda=-1.58^{+0.14}_{-0.22}$) is typical for an unobscured quasar \citep{VB01}.
Figure~\ref{fig:J0859_zoom} shows the continuum of J0859+0022 with its best-fit power-law model.
Second, the presence of the iron emission line forest further supports the identification of this quasar as a type-I quasar (Figure~\ref{fig:MgII}).
Indeed, if we fit the continuum of J0859+0022 with a power-law model only, the reduced $\chi^2$ of the data in the iron fitting window is worse by $\Delta \chi_\nu^2\approx0.3$ than the best-fit power-law + iron model,
though the iron is unambiguously detected.
Therefore, J0859+0022 is likely a low-luminosity type-I AGN and the high-redshift analogue of the local Narrow Line Seyfert 1 galaxies \citep[NLS1;][]{Osterbrock85, Constantin03}.
Indeed, J0859+0022 has similar continuum and emission line properties to the $2<z<4.3$ NLS1s reported in a spectroscopic search of high-redshift type-II quasar candidates \citep{Alexandroff13}.

%

%


\section{BH mass and Eddington Ratio} \label{sec:BHmass}
Emission lines of type-I quasars are broad because of the Doppler motion in the broad-line region (BLR) gas.
We can use the width of these lines to estimate the central black hole mass,
with the assumption that the BLR gas is gravitationally bound to the SMBHs.
The virial mass is given by:
\begin{equation}
M_\mathrm{BH}=f G^{-1}  v^2_\mathrm{BLR} R_\mathrm{BLR},
\end{equation}
where $G$ is the gravitational constant, $v_\mathrm{BLR}$ is the velocity of the orbiting BLR gas around the central black hole, and  $R_\mathrm{BLR}$ is the distance from the black hole to the BLR.
The scaling factor $f$ takes into account the geometry.

For the Mg{\sc ii}-based BH mass estimate of the HSC quasars, we follow the scaling relation (the so-called ``single-epoch" method) given in \citet{Vestergaard09}:
\begin{equation}
M_\mathrm{BH} = 10^{6.86} \left(\frac{\mathrm{FWHM(Mg{\sc II})}}{10^3\ \mathrm{km\ s^{-1}}}\right)^2 \left(\frac{\lambda L_\lambda\ (3000\mathrm{\AA})}{10^{44}\ \mathrm{erg\ s^{-1}}}\right)^{0.5}M_\odot,\label{eq:VO09}
\end{equation}
where FWHM(Mg{\sc ii}) is the full width at half maximum of the Mg{\sc ii} line and $\lambda L_\lambda\ (3000\mathrm{\AA})$ is the monochromatic luminosity at rest-frame $3000$\AA.
This relation is based on reverberation mapping studies which found a tight relation between continuum luminosity and BLR radius at low redshift \citep[e.g.,][]{Kaspi05}.
The measurement uncertainties of the virial black hole masses are derived by propagating the measurement errors of the Mg{\sc ii} line widths and the monochromatic luminosity.
The systematic uncertainty is usually larger than the errors of the single-epoch masses.
We adopt $0.5$ dex as a $1\sigma$ uncertainty of the virial mass estimate \citep{Shen13_review}.
It is possible that part of the narrow Mg{\sc ii} emission line of J0859+0022 originates from the narrow-line region; however, we use the observed width as a ``broad" line because the X-Shooter spectrum cannot clearly distinguish the two components.
Therefore, our line measurement may overestimate the line flux and underestimate the line width of the BLR gas.

We also measure the C{\sc iv}-based mass for three objects (J0859+0022, J1152+0055, J2239+0207), 
without absorption lines at the C{\sc iv} line peaks.
We use the scaling relation of \citet{Vestergaard06}:
\begin{equation}
M_\mathrm{BH} = 10^{6.66} \left(\frac{\mathrm{FWHM(C{\sc IV})}}{10^3\ \mathrm{km\ s^{-1}}}\right)^2 \left(\frac{\lambda L_\lambda(1350\mathrm{\AA})}{10^{44}\ \mathrm{erg\ s^{-1}}}\right)^{0.53}M_\odot .\label{eq:V06}
\end{equation}
We assume the same systematic uncertainty ($0.5$ dex) for the C{\sc iv}-based mass as the Mg{\sc ii}-based mass \citep{Shen13_review}.

Once we have measured the black hole mass we can determine the Eddington luminosity of the quasars.
The Eddington luminosity $L_\mathrm{Edd}$ is proportional to the black hole mass:
\begin{equation}
L_\mathrm{Edd}=1.3\times10^{38} \left(\frac{M_\mathrm{BH}}{M_\odot} \right)\ \mathrm{erg\ s^{-1}}.
\end{equation}
For each quasar, we compute the bolometric luminosity $L_\mathrm{bol}$ to determine the Eddington ratios.
For this purpose, we apply the bolometric correction given in \citet{Richards06a}:
\begin{equation}
L_\mathrm{bol}= 5.15\ \lambda L_\lambda (3000\mathrm{\AA})\ \mathrm{erg\ s^{-1}}.\label{eq:BC}
\end{equation}

\begin{figure}[tb!]
\plotone{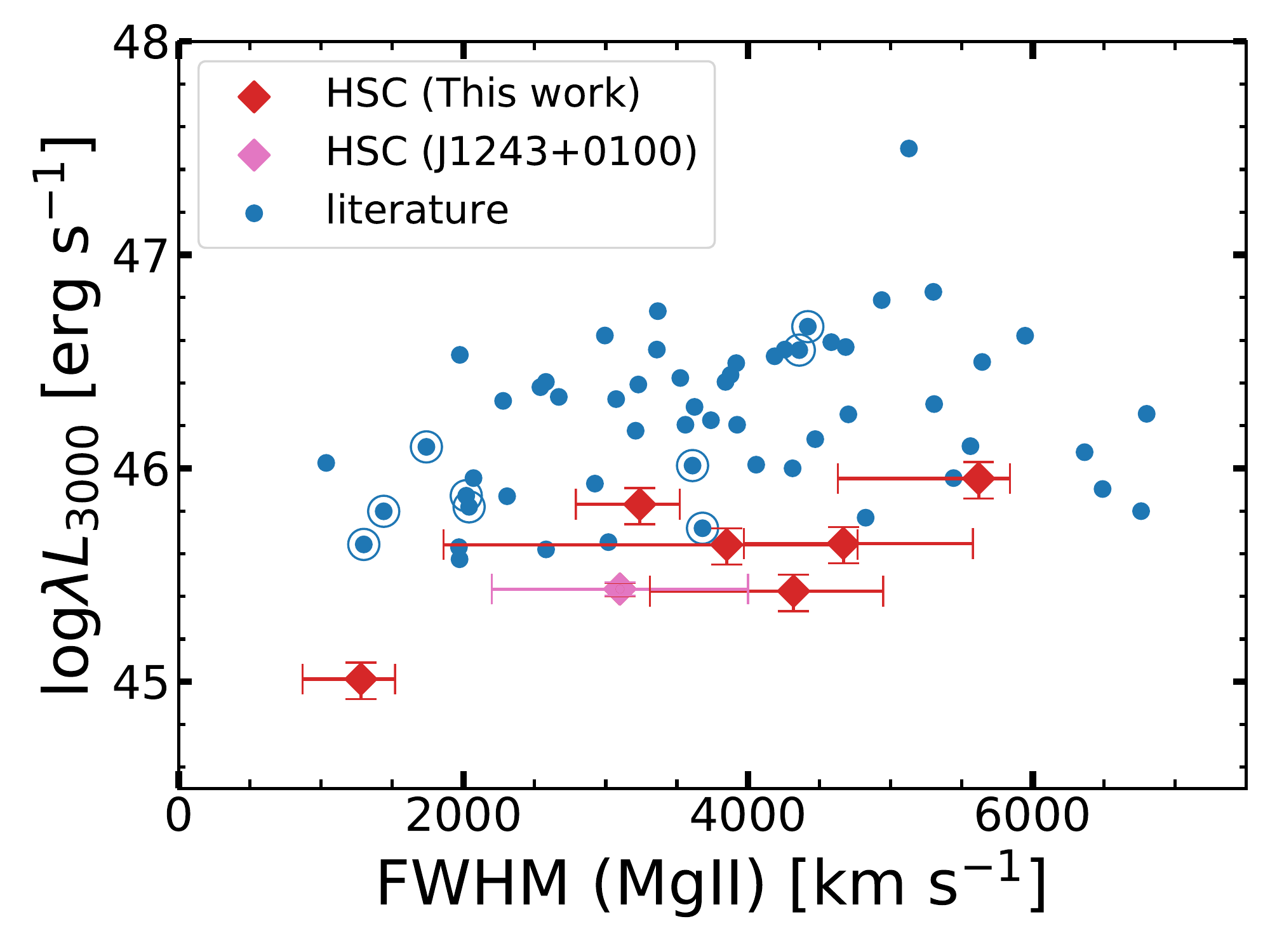}
\caption{Distribution of Mg{\sc ii} FWHM and monochromatic luminosity at rest-frame $3000$\AA\ for $z>5.8$ quasars. 
The HSC quasars in this work are shown as filled red diamonds, while HSCJ1243+0100 at $z=7.07$  \citep{Matsuoka19} is shown as a magenta diamond.
Other $z>5.8$ quasars whose $M_\mathrm{BH}$ measurements have been derived in the literature are shown as blue dots.
The luminosity of the non-HSC quasars is measured with the same cosmology as that of the HSC quasars.
The CFHQS quasars from \citet{Willott10a} are marked with circles.
}.\label{fig:MgIIline}
\end{figure}

We compile the Mg{\sc ii} line measurements of $z>5.8$ quasars from the literature to compare with the SMBH properties of the HSC quasars.
Mg{\sc ii} line measurements have been made for $60$ quasars in total \citep{Kurk07, Jiang07, Willott10a, Mortlock11, DeRosa11,  DeRosa14, Wu15, Mazzucchelli17, Banados18, Eilers18, Wang18, Shen19, Fan19, Tang19, Matsuoka19}. 
We use the most recent measurements when more than one measurement exists for the same objects, while for CFHQS quasars we use the original measurements by W10\footnote{\citet{Shen19} measured Mg{\sc ii}-based masses for 3 CFHQS quasars: J0050+3445 ($z=6.25$), J0055+0146 ($z=5.95$), and J0221-0802 ($z=6.20$).}.
Specifically, the five $z\sim6$ quasars in \citet{Kurk07} are replaced with measurements by \citet{Jiang07} and \citet{DeRosa11}.
The $z>6.5$ quasars in \citet{Mortlock11} and \citet{DeRosa14} are replaced with measurements by \citet{Mazzucchelli17}.
Figure~\ref{fig:MgIIline} shows the distribution of the measured Mg{\sc ii} line FWHM and the $3000$\AA\ monochromatic luminosity for this entire sample, including the HSC quasars in this paper and HSC J1243+0100 \citep{Matsuoka19}.

\begin{figure*}[ht!]
\plotone{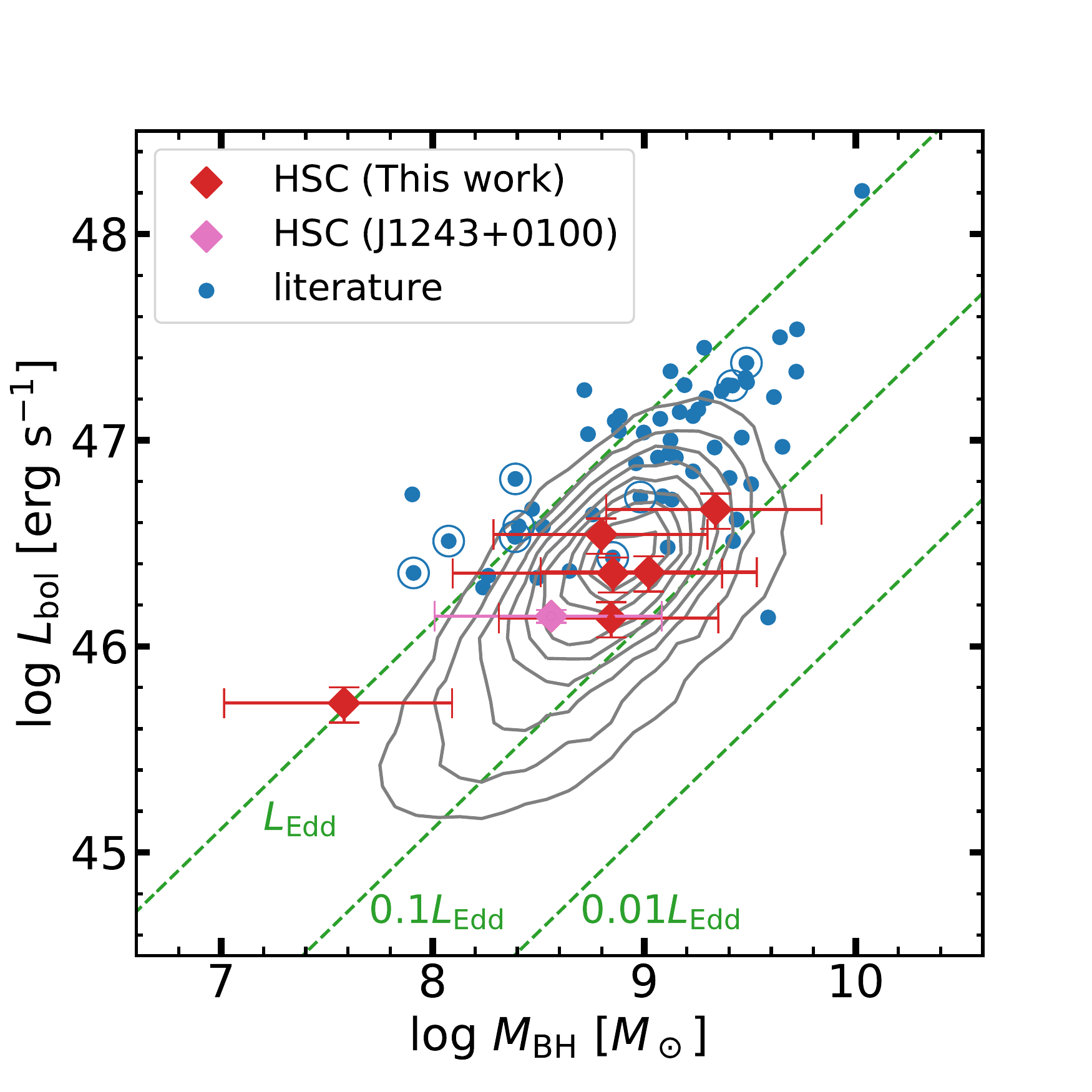}
\caption{The SMBH mass-luminosity plane of $z>5.8$ quasars, the black hole mass $M_\mathrm{BH}$ of which have been measured to date with Mg{\sc ii}.
The symbols and colors are the same as Figure~\ref{fig:MgIIline}.
The quoted virial masses are derived using the scaling relation of  \citet{Vestergaard09}.
For non-HSC quasars, we quote their Mg{\sc ii} line and \replaced{$3000$\AA\ monochromatic  luminosity}{continuum} measurements to calculate $M_\mathrm{BH}$  and bolometric luminosity with the same relation and cosmology as those applied for the HSC quasars (Eq.~\ref{eq:VO09}).
The systematic uncertainty of the virial black hole mass measurement ($0.5$ dex; \citealt{Shen13_review}) is included in the error bars.
Contours show the distribution of the $z\sim2$ SDSS DR7 quasars  \citep{Shen11}.
The diagonal lines show Eddington luminosities of $L_\mathrm{bol}/L_\mathrm{Edd}=1, 0.1, 0.01$ from top left to bottom right.
}.\label{fig:MBH}
\end{figure*}

Our measurements of the virial black hole masses and the Mg{\sc ii}-based Eddington ratios are summarized in Table~\ref{tab:shellqs_BHmass}.
Figure~\ref{fig:MBH} shows the distribution of the HSC quasars in the SMBH mass-luminosity plane.
We include the $0.5$ dex systematic uncertainty of the black hole mass in the error bars, which dominates the uncertainty of our mass measurements.
We measured the black hole masses of the non-HSC quasars with the same scaling relation (Equation~\ref{eq:VO09}) and cosmology adopted in this work,
rather than just quoting the values in the reference papers.
Their bolometric luminosities are also calculated from the $3000$\AA\  monochromatic luminosity in the same way as the HSC quasars  (Equation~\ref{eq:BC}).
We find that the Mg{\sc ii}-based $M_\mathrm{BH}$ of the six HSC quasars span a wide range of $3.8\times10^7 M_\odot\leq M_\mathrm{BH} \leq 2.2\times10^9M_\odot$ and accordingly a wide range of the Eddington ratio of $0.16\leq L_\mathrm{bol}/L_\mathrm{Edd}\leq 1.1$.
J0859+0022 hosts a $10^7M_\odot$ SMBH, which is the least massive SMBH known at $z>5.8$, and has a similarly high Eddington ratio to the most active quasars at the same epoch.
On the other hand, it is remarkable that the other HSC quasars 
are powered by $10^9M_\odot$ SMBHs with sub-Eddington accretion, as does HSC J1243+0100 \citep[$M_\mathrm{BH}=(3.3\pm2.0)\times10^8M_\odot$, $L_\mathrm{bol}/L_\mathrm{Edd}=0.34\pm0.20$;][]{Matsuoka19}.
The average Eddington ratio of those six sub-Eddington quasars (i.e., excluding J0859+0022) is $\left<L_\mathrm{bol}/L_\mathrm{Edd}\right>=0.24\pm0.10$.
Their masses and accretion rates overlap with the range of typical $z\sim2$ SDSS DR7 quasars \citep{Shen11}.
Note that the quoted value for J1205-0000, the dust-reddened quasar, is not corrected for extinction; therefore the quoted mass and accretion rate in Table~\ref{tab:shellqs_BHmass} should be considered to be upper and lower limits, respectively.

\begin{deluxetable*}{lCCCCCC}[bt!]
\tablecaption{Black Hole Mass and Eddington Ratio \label{tab:shellqs_BHmass}
}
\tablecolumns{7}
\tablenum{4}
\tablewidth{0pt}
\tablehead{
\colhead{} &
\colhead{J1205-0000} &
\colhead{J0859+0022} &
\colhead{J1152+0055} &
\colhead{J2239+0207} &
\colhead{J1208-0200} &
\colhead{J2216-0016} 
}
\startdata
$\lambda L_{1350}$ [$10^{45}$ erg s$^{-1}$] & $\cdots$& $1.63\pm0.09$ & $10.0\pm0.1$ & $6.45\pm0.20$ & $\cdots$ & $\cdots$  \\
$\lambda L_{3000}$ [$10^{45}$ erg s$^{-1}$] & $8.96\pm0.66$& $1.03\pm0.10$ & $6.77\pm0.11$ & $4.44\pm0.08$ & $4.38\pm0.04$ & $2.66\pm0.05$  \\
$M_\mathrm{BH}$(C{\sc iv}) [$10^8  M_\odot$] & $\cdots$& 0.34_{-0.02}^{+0.04}  & 14.1_{-1.4}^{+1.4}& 8.9_{-3.4}^{+2.8} & $\cdots$ & $\cdots$ \\
$M_\mathrm{BH}$(C{\sc iv}, cor) [$10^8  M_\odot$] & $\cdots$ & 0.14_{-0.01}^{+0.02}  & 11.8_{-1.2}^{+0.9} & 6.3_{-2.5}^{+2.0} & $\cdots$ & $\cdots$ \\
$M_\mathrm{BH}$(Mg{\sc ii}) [$10^8  M_\odot$] & 22_{-6}^{+2} & 0.38_{-0.18}^{+0.10}  & 6.3_{-1.2}^{+0.8}& 11_{-2}^{+3} & 7.1_{-5.2}^{+2.4} &7.0_{-2.3}^{+1.4} \\
$L_\mathrm{bol}/L_\mathrm{Edd}$ & $0.16^{+0.04}_{-0.02}$ & 1.1^{+0.5}_{-0.3}  & 0.43^{+0.08}_{-0.05}& $0.17^{+0.04}_{-0.05}$ & $0.24^{+0.18}_{-0.08}$ & $0.24^{+0.06}_{-0.01}$ \\
\enddata
\tablecomments{
The black hole mass errors quoted in this table are measurement errors, while there is an additional systematic uncertainty of 0.5 dex in the single-epoch mass measurement \citep{Shen13_review}.
The Eddington ratio is based on the Mg{\sc ii}-based $M_\mathrm{BH}$ \citep{Vestergaard09}. 
Two C{\sc iv}-based $M_\mathrm{BH}$ measurements are quoted: 
$M_\mathrm{BH}$(C{\sc iv}) is the virial mass based on \citet[][Eq.~\ref{eq:V06}]{Vestergaard06}, while $M_\mathrm{BH}$(C{\sc iv}, cor) is the virial mass based on \citet{Coatman17}, which takes into account the C{\sc iv} blueshift with respect to the systemic redshift.
}
 \end{deluxetable*}


This large range in the SMBH properties of $z\gtrsim6$ quasars has been seen in other recent papers.
All known luminous quasars at $z>7.0$ have accretion rates around the Eddington limit (i.e., $L_\mathrm{bol}/L_\mathrm{Edd}\sim1$; \citealt{Mortlock11, Banados18, Wang18}).
Lower-luminosity quasars in W10 show similarly high SMBH activity.
On the other hand, \citet{Mazzucchelli17} and \citet{Shen19} found that a significant fraction of luminous $z\gtrsim6$ quasars host sub-Eddington SMBHs (i.e., $L_\mathrm{bol}/L_\mathrm{Edd}\sim0.1$) and the entire Eddington ratio distribution is not significantly different from that of the lower-redshift quasars at the same luminosity range.
The SMBH activity of the HSC quasars agrees somewhat with the conclusion of \citet{Mazzucchelli17} and \citet{Shen19}, albeit with our small sample size.
Note that \citet{KimY18} identify another sub-Eddington SMBH at $z=5.9$ based on a C{\sc iv}-based mass measurement.

\begin{figure}[t!]
 \plotone{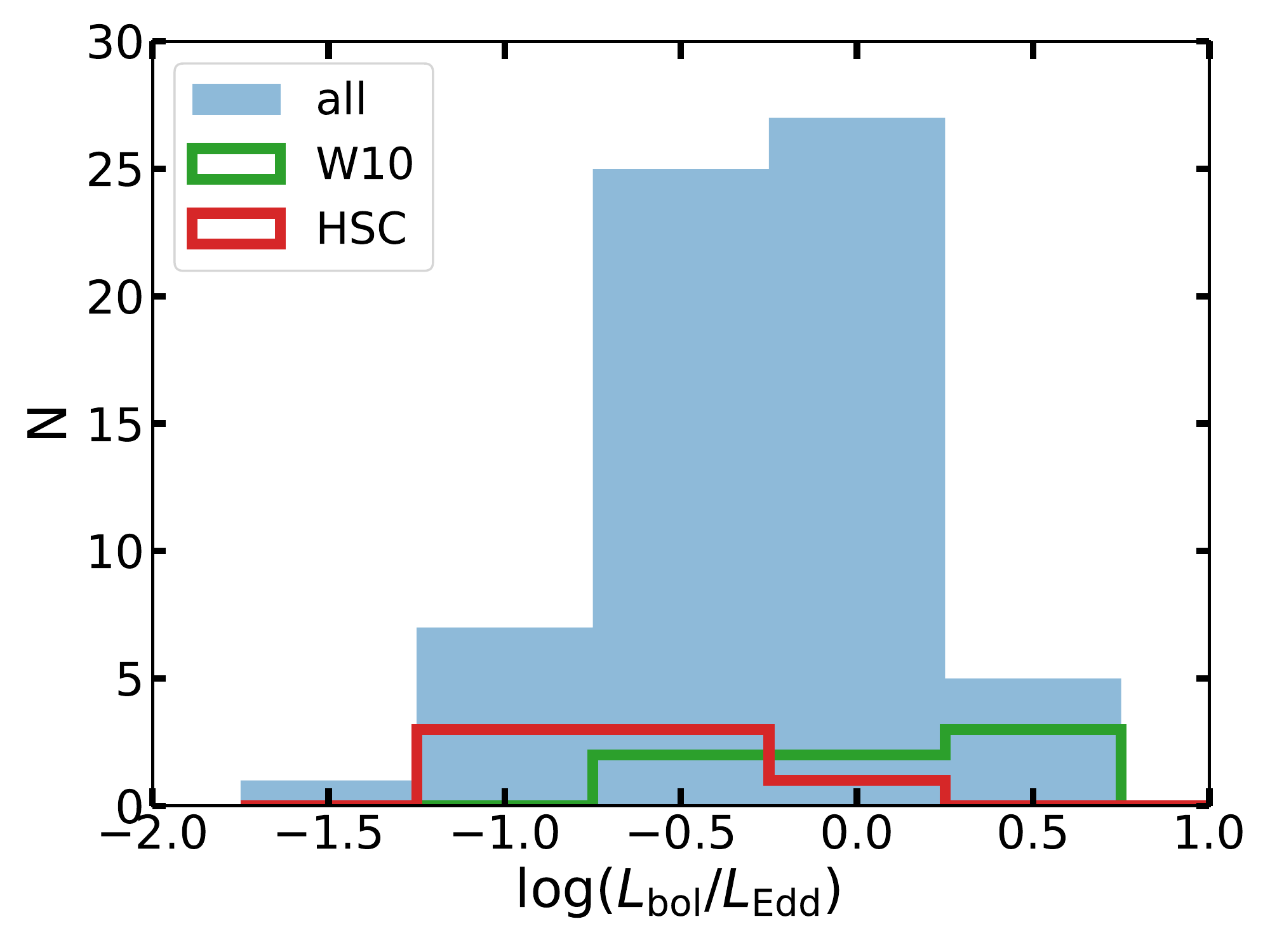}
 \caption{Histogram of the Eddington ratio for $z>5.8$ quasars based on the Mg{\sc ii}-based scaling relation of \citet[][Equation~\ref{eq:VO09}]{Vestergaard09}.
 The bin step is $0.5$ dex, which is the same as the systematic uncertainty \citep{Shen13_review}.
The HSC and CFHQS quasars at $L_\mathrm{bol}<10^{47} \mathrm{erg\ s^{-1}}$ are shown in red and green, respectively.
 The shaded histogram shows all of the $z>5.8$ quasars whose Mg{\sc ii} lines have been measured.
}\label{fig:MBH_Eddratio}
 \end{figure}

While W10 use the same scaling relation that we used \citep{Vestergaard09} with a slightly larger bolometric correction factor of $6.0$,
most of the CFHQS quasars at $L_\mathrm{bol}<10^{47}\ \mathrm{erg\ s^{-1}}$ are still around or above the Eddington limit in Figure~\ref{fig:MBH}.
Figure~\ref{fig:MBH_Eddratio} shows the histogram of the Eddington ratio for the HSC quasars (including J1243+0100) and the CFHQS quasars from W10.
\citet{Shen19} argue that the apparent difference between the Eddington ratio distribution of their objects and W10 is partly due to the different Mg{\sc ii} recipe used for mass estimate and different bolometric correction.
However, most of the CFHQS quasars still have luminosities near or over the Eddington limit, even when we re-calculate their Mg{\sc ii}-based $M_\mathrm{BH}$ using the same calibration as the HSC quasars.
This result may suggest that the global Eddington ratio distribution is much broader than measured for the CFHQS and HSC quasars, taking into account the small sample sizes ($<10$) in both studies.
To explore this in detail will require a sample of low-luminosity quasars comparable in size to the high-luminosity sample.

\begin{figure}[t!]
 \plotone{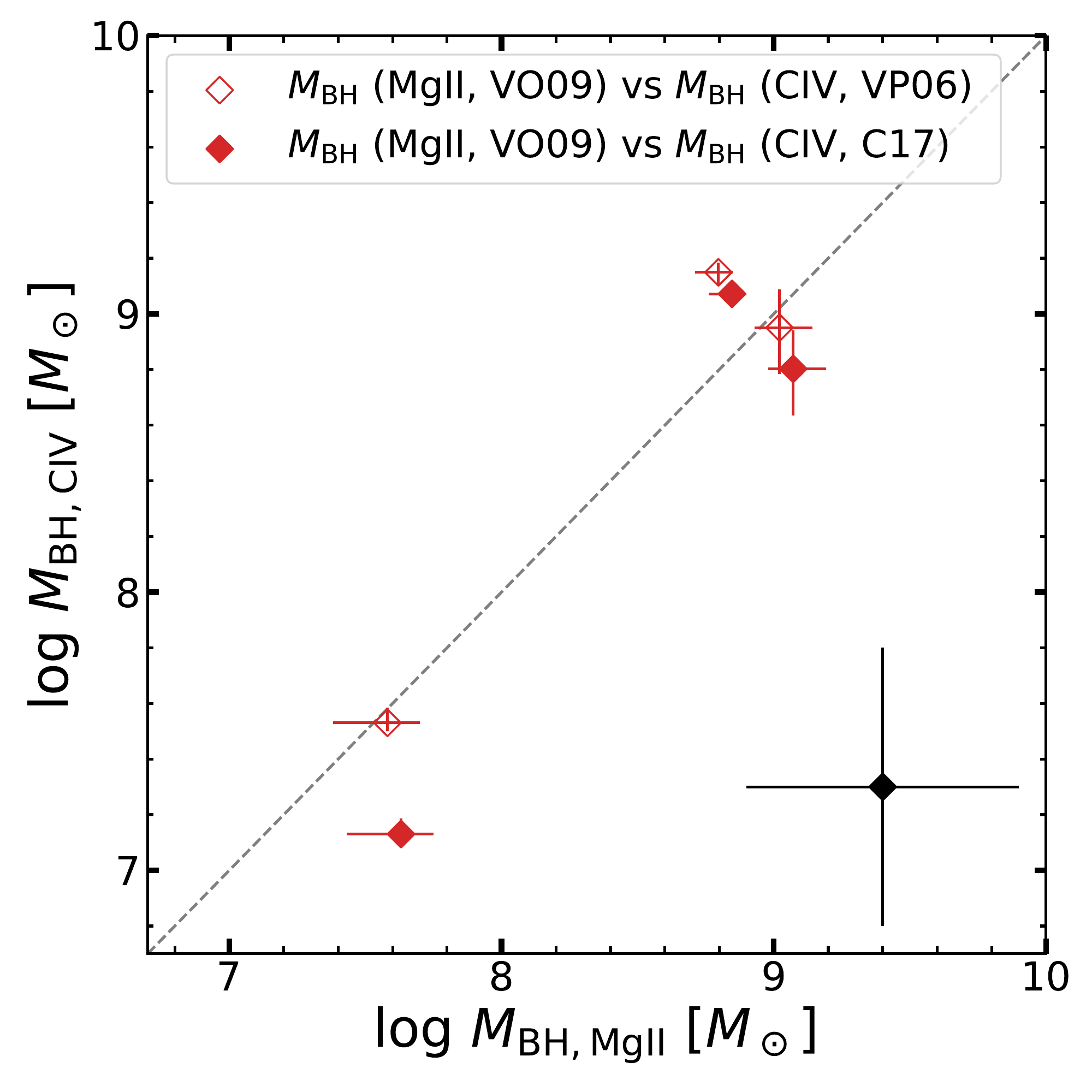}
 \caption{Comparison of the Mg{\sc ii}-based and C{\sc iv}-based black hole masses for the HSC quasars whose masses have been measured with both emission lines (J0859+0022, J1152+0055, and J2239+0207).
 The Mg{\sc ii}-based mass shown in this figure is based on the calibration of \citet{Vestergaard09}.
 We show the C{\sc iv}-based mass based on \citet{Vestergaard06} with open symbols and  \citet{Coatman17} with filled symbols.
 The latter calibration takes into account the C{\sc iv} blueshift with respect to the systemic redshift.
 The systematic uncertainties of the mass measurements ($0.5$ dex) are shown in the black point in lower right.
 }\label{fig:MBH_comp}
 \end{figure}

We investigate how the derived virial mass depends on the choice of mass scaling relations and mass estimators.
We measure the Mg{\sc ii}-based SMBH mass for the six HSC quasars in this work with the calibration of \citet{Shen11}.
The Mg{\sc ii} recipe of \citet{Shen11} gives systematically higher black hole masses at the luminous end than \citet{Vestergaard09}, which we used in this work.
For example, the most luminous $z\sim6$ quasar konwn, SDSS J1030+0100 \citep{Wu15}, shifts to higher mass by a factor of two ($0.3$ dex), which accordingly reduces the Eddington ratio by the same factor.
However, its effect is tiny ($<0.15$ dex) at the luminosity range of the HSC quasars ($L_\mathrm{bol}\lesssim10^{47}$ erg s$^{-1}$), when compared to the 0.5 dex systematic uncertainty of the virial mass measurements.
We also measure the C{\sc iv}-based black hole mass for three HSC quasars (J0859+0022, J1152+0055, J2239+0207), the C{\sc iv} emission lines of which are not severely affected by self-absorption at the line centers.
The results are also reported in Table~\ref{tab:shellqs_BHmass}.
Figure~\ref{fig:MBH_comp} compares the single-epoch mass measurements based on  Mg{\sc ii} and C{\sc iv}.
The C{\sc iv}-based black hole masses of the three quasars are in agreement with the Mg{\sc ii}-based mass within $1\sigma$, when the measurement and systematic uncertainties are taken into account \citep{Shen13_review}.
Therefore, the observed SMBH activity of the three HSC quasars is robust to the choice of the mass estimator.

In addition, we leverage an empirical correction of the C{\sc iv}-based mass \citep{Coatman17} using C{\sc iv} blueshifts with respect to systemic redshifts.
This correction is calibrated with luminous quasars at $1.5<z<4.0$, which typically show $\approx1000-5000$ km s$^{-1}$ C{\sc iv} blueshifts.
The corrected C{\sc iv}-based black hole mass is also reported in Table~\ref{tab:shellqs_BHmass} and Figure~\ref{fig:MBH_comp}.
This gives only a slightly better agreement with the Mg{\sc ii}-based mass for J1152+0055, which has a blueshift of $\Delta v_\mathrm{C{\sc IV}-[CII]}=590^{+130}_{-190}$ km s$^{-1}$.
On the other hand, the deviation from the Mg{\sc ii}-based mass gets bigger for J0859+0022 ($0.39$ dex), while still in the $1\sigma$ systematic uncertainty.
This is likely because the calibration is not sensitive to the small C{\sc iv} velocity offset of J0859+0022 ($60$ km s$^{-1}$ {\it redshift}); most quasars in \citet{Coatman17} have $\gtrsim1000$ km s$^{-1}$ C{\sc iv} blueshifts.
It may also be possible that the observed C{\sc iv} line of J0859+0022 is affected by the BAL absorption in the blue wing (Figure~\ref{fig:CIV}), which makes the line width narrower than the intrinsic shape, and causes us to underestimate the black hole mass.

\section{Implications on the Early SMBH Growth}\label{sec:MBHacc}

\begin{figure*}[htb!]
\plotone{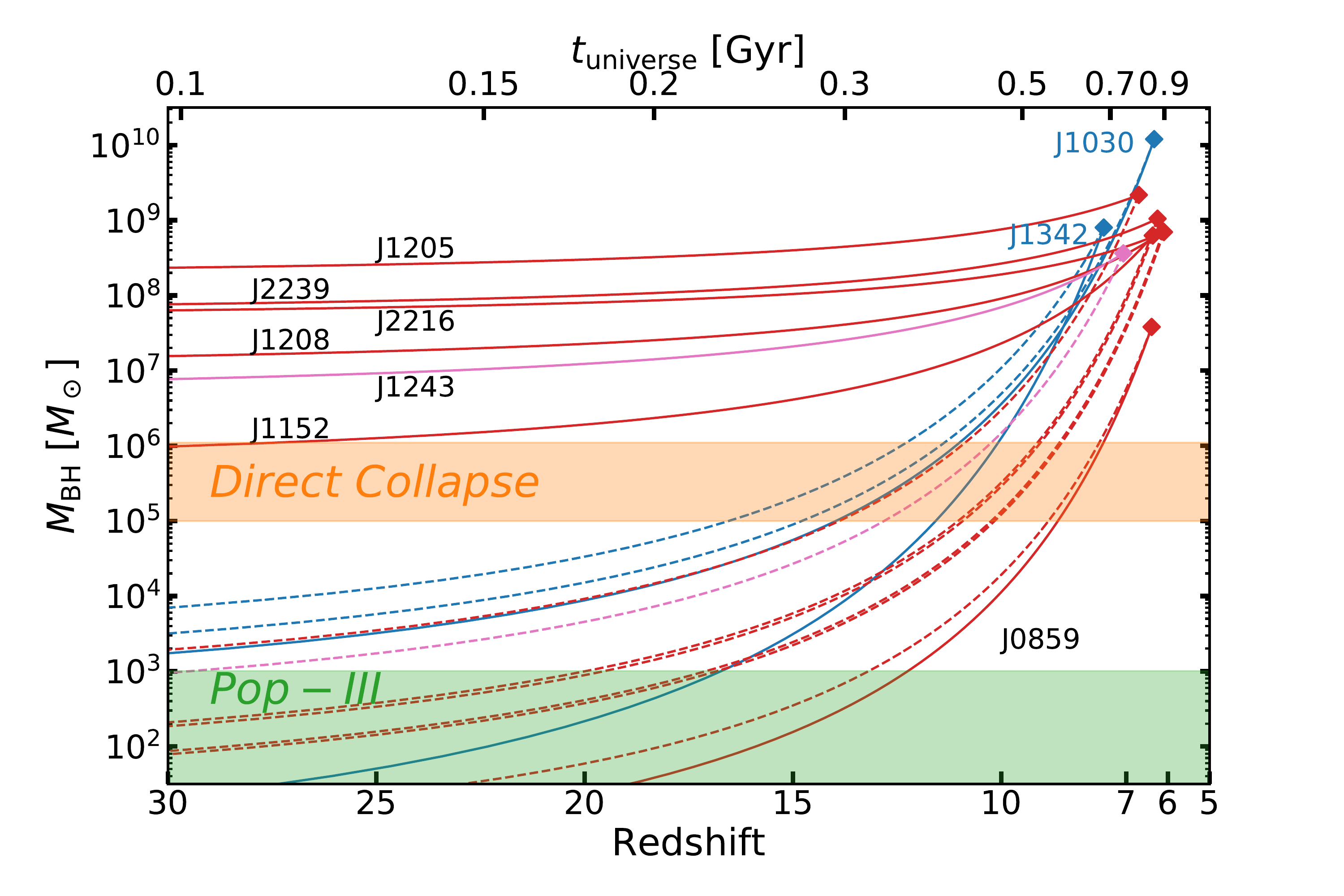}
\caption{Estimated growth history of the $z\geq6.1$ HSC quasars.
The $x$-axis is given in both redshift (bottom) and the time since the Big Bang (top).
Solid lines show the case where SMBHs grow at the observed Eddington ratio at all times, while dashed lines show the case where SMBHs grow at the Eddington limit $L_\mathrm{bol}/L_\mathrm{Edd}=1$.
The six HSC quasars in this work are shown as red, while HSC J1243+0100 is shown as magenta.
The observed masses are shown as filled diamonds.
For comparison,
two luminous $z>6$ quasars are also shown in blue;
J1342+0928 at $z=7.54$ from \citet{Banados18} and J0100+2802 at $z=6.33$ from \citet{Wu15}.
The shaded regions correspond to the mass ranges of Pop-III remnant black holes ($M_\mathrm{seed}\lesssim10^3M_\odot$; green) and direct collapse black holes ($M_\mathrm{seed}\sim10^{5-6}M_\odot$; orange).
}\label{fig:MBH_growth}
\end{figure*}

The growth of a SMBH is exponential if it keeps a constant Eddington ratio.
The timescale for a seed black hole ($M_\mathrm{seed}$) to reach a given $M_\mathrm{BH}$ is
\begin{equation}
t_\mathrm{grow}=\tau \ln{\left(\frac{M_\mathrm{BH}}{M_\mathrm{seed}}\right)} ,
\end{equation}
where $\tau$ is the $e$-folding timescale of 
\begin{equation}
\tau=0.45 \left(\frac{\eta}{1-\eta}\right)\left(\frac{L_\mathrm{bol}}{L_\mathrm{Edd}}\right)^{-1} \ \mathrm{Gyr}.\label{eq:tgrow}
\end{equation}
The radiation efficiency $\eta$ is the factor of how efficiently the accreting mass is converted to radiation.
Figure~\ref{fig:MBH_growth} shows the estimated growth history of the HSC quasars.
We trace back to $z=30$ ($10^8\ \mathrm{yr}$ after the Big Bang) when the first stars and galaxies were thought to have formed in the universe \citep{Bromm04, Bromm11}.
The radiation efficiency is assumed to be $\eta=0.1$ (i.e., a standard thin accretion disk; \citealt{Shakura76}).
Two scenarios are considered here: the first assumes the Eddington ratios we derived from our observations  (Section~\ref{sec:BHmass}), and the second assumes the Eddington limit $L_\mathrm{bol}/L_\mathrm{Edd}=1$.
For comparison, we also show in Figure~\ref{fig:MBH_growth} the growth paths of 
J1342+0928 at $z=7.54$ \citep{Banados18}, the highest-redshift quasar known to date, and J0100+2802 at $z=6.33$ \citep{Wu15}, the most luminous $z>6$ quasar to date.

Among the seven HSC quasars, J0859+0022 reaches the mass range of the Pop-III remnants  at $z\gtrsim10$ in both scenarios above ($M_\mathrm{seed}\lesssim10^3M_\odot$; \citealt{Hirano14}) due to its relatively low mass and high Eddington ratio.
This is different from the cases of J1342+0928 and J0100+2802, where the seed black holes need to form earlier or need to be more massive when they form.
On the contrary, the initial growth of the other HSC quasars with sub-Eddington ratios require $M_\mathrm{seed}\gtrsim10^6M_\odot$ seed black holes, if we assume the current growth speeds.
This is even more massive than what the direct collapse model predicts ($M_\mathrm{seed}\sim10^{5-6}M_\odot$; \citealt{Latif16}).
There is currently no plausible formation model for such an extremely heavy seed black hole.
Note that there is a large systematic uncertainty of the Eddington ratio for each  object (Figure~\ref{fig:MBH});
however, our conclusion that the bulk of the HSC quasars  indeed host sub-Eddington SMBHs is robust, and thus this challenge to seed black hole formation models remains.

Figure~\ref{fig:MBH_growth} also suggests that their seed masses do fall into the range of the Pop-III remnants if we assume Eddington limit accretion.
Those SMBHs are likely to be in a quiescent phase, the mass accretion must have been longer 
at earlier epochs, but switched to a less active mode by the time we observe them.
It is still possible that those sub-Eddington SMBHs will switch to more active mode again if they accrete large amount of cold gas through major mergers.
However, our ALMA observations of the HSC quasars in this paper (all but J1205-0000 and J1243+0100) have revealed that they do not have [C{\sc ii}]-bright companion galaxies \citep{Izumi18, Izumi19}, as some luminous quasars do \citep{Decarli17}.
Therefore, it is likely that the sub-Eddington SMBHs in this work will not grow much more rapidly at least over the next $100$ Myr (i.e., typical quasar lifetime; \citealt{Soltan82}).
Such quiescent SMBHs have been found at $z\lesssim5$ at similar luminosity \citep{Trakhtenbrot16, Ikeda17}.
We stress that the importance of the HSC quasars in this study is in showing the quenching of efficient SMBH growth at earlier epoch than those studies.
We may be witnessing the first ``down-sizing" of the most massive SMBHs in the context of the anti-hierarchical SMBH evolution \citep[e.g.,][]{Ueda03, Ueda14}.

The relatively low SMBH activity of the HSC quasars implies that the mass accretion at constant Eddington ratio (as assumed in Equation~\ref{eq:tgrow}) can no longer explain the SMBH growth at $z\gtrsim6$, and that one should consider the time evolution of Eddington ratios from seed black holes.
It is relatively easy for heavy seed black holes ($M_\mathrm{seed}\sim10^{5-6}M_\odot$) to reach $10^9M_\odot$ in the early universe, if part of their accretion is at the Eddington limit. 
However, the formation of such a seed needs critical conditions which would not be met for all SMBHs in the universe \citep[e.g.,][]{Latif16, Chon16, Hirano17}.
Moreover, \citet{Shirakata16} show that the local  bulge-to-BH mass ratio cannot be reproduced in their semi-analytical model, if the seed black holes are all $M_\mathrm{seed}=10^5M_\odot$.

In this context, another plausible scenario would be that early SMBHs experience episodic super-Eddington growth.
\citet{Collin04} argue that SMBH growth is not capped at the Eddington limit, but is regulated by the supply of infalling mass (i.e., the maximum accretion rate is irrespective of the SMBH mass).
Theoretical studies suggest that super-Eddington accretion can be preferentially achieved in metal-poor environments ($Z<10^{-2}Z_\odot$), where the nuclear feedback is less efficient in regulating mass accretion than in the current universe \citep{Inayoshi16, Toyouchi19}.
There has been no clear examples of such a super-Eddington SMBH at $z\gtrsim6$ to date, however
the discovery of a mildly obscured quasar at $z=6.699$ (J1205-0000; Section~\ref{sec:J1205}) may support this scenario because the mass accretion onto SMBHs would be  most efficient during the dusty phase after starburst of their hosts \citep{Hopkins08, Kim15}.
J1205-0000 is perhaps in the transitional phase from a dusty quasar to a type-I quasar just after experiencing intense mass growth associated with dusty major mergers of its host galaxy \citep{Decarli17}.
Observational evidence of high-redshift super-Eddington SMBHs ($L_\mathrm{bol}/L_\mathrm{Edd}\gg1$), perhaps hosted by dusty starburst galaxies, is required to test the episodic SMBH growth scenario.

Finally, the variety seen in the SMBH properties of $z>5.8$ quasars requires us to revisit the redshift evolution of the black hole mass function (BHMF).
W10 show the first constraint on the $z\sim6$ BHMF  by interpreting the CFHQS quasar luminosity function of \citet{Willott10b} in the context of their $M_\mathrm{BH}$ measurements.
On the other hand, \citet{Matsuoka18c} show the $z\sim6$ luminosity function with a compilation of $110$ quasars spanning a wide luminosity range, showing
a flatter power-law slope $\alpha$ at the faint end than \citet{Willott10b} ($\alpha=-1.23$, where W10 quoted $\alpha=-1.5$).
Taken together with the wider Eddington ratio distribution that we showed in Figure~\ref{fig:MBH_Eddratio}, 
the studies of HSC quasars suggest that 
the slope of the $z\sim6$ BHMF is flatter than that of W10.
The flat low-mass end slope is in line with the scenario
in which the seed black holes can easily grow to large masses through a ``fast track" from seed black holes.
Given that we have $M_\mathrm{BH}$ measurements of only seven HSC quasars, it is necessary to construct a larger mass sample to better understand the early SMBH growth with better statistics.
In addition, our current near-infrared observations are limited by the sensitivity of the ground-based 8m telescopes.
Future facilities such as the James Webb Space Telescope \citep{Gardner06} and next-generation large ground-based telescopes will reveal the SMBH properties of even lower mass or less active SMBHs.


\section{Summary} \label{sec:summmary}

This paper is the sixth paper of the SHELLQs project, a large optical survey of low-luminosity quasars in the reionization epoch ($z>5.7$) with the HSC-SSP. 
We have presented near-infrared (and some optical) observations of six quasars at $6.1\leq z\leq6.7$ selected from the first two discovery papers \citep{Matsuoka16, Matsuoka18a}.
Their absolute continuum magnitudes ($M_{1450}$) are
among the faintest of $z>5.8$ quasars whose central black hole masses have been measured.

We observed three quasars each with VLT/X-Shooter and Gemini-N/GNIRS to measure their SMBH mass and accretion rate with the single-epoch method.
The broad emission lines of Mg{\sc ii} $\lambda2798$  and C{\sc iv} $\lambda1549$ were detected in all targets, as were underlying continuum and other emission lines such as C{\sc iii]} $\lambda1909$ and Si{\sc iv} $\lambda1397$ in some cases.
The near-infrared spectra are fitted with power-law continuum, iron emission line templates, and Gaussian profiles for emission lines.
The velocity shifts of C{\sc iv} emission lines ($\lesssim 400-600$ km s$^{-1}$) are broadly in agreement with luminous $z\sim6$ quasars.  
No large Mg{\sc ii} blueshifts ($\gtrsim1000$ km s$^{-1}$) are observed for the five quasars which have atmoic [C{\sc ii}] $158\micron$ redshifts. 
We find that one quasar, J1205-0000 at $z_\mathrm{MgII}=6.699^{+0.007}_{-0.001}$ is a modestly obscured quasar associated with (at least) two strong BAL troughs in the blue wings of C{\sc iv} and N{\sc v} $\lambda1240$.
Another quasar, J0859+0022 at $z_\mathrm{[CII]}=6.3903$, has remarkably strong and narrow emission lines ($\mathrm{FWHM=1200-1400\ km\ s^{-1}}$),
which we classify as a high-redshift analogue of a narrow line Seyfert~1 galaxy rather than a type-II quasar, 
from the fact that it has a blue continuum and iron emission line forest.

The Mg{\sc ii}-based single-epoch mass measurements of the six HSC quasars reveal a variety of SMBH properties, from rapidly growing SMBHs to modestly accreting massive SMBHs, with masses of 
 $3.8\times10^{7} M_\odot\leq M_\mathrm{BH} \leq 2.2\times 10^{9} M_\odot$ and  Eddington ratios of $0.16\leq L_\mathrm{bol}/L_\mathrm{Edd}\leq 1.1$.
 It is remarkable that the majority hosts $M_\mathrm{BH}\sim10^9 M_\odot$ SMBHs with sub-Eddington accretion.
 The observed Eddington ratio distribution of the HSC quasars shifts to lower accretion rates than the CFHQS quasars in \citet{Willott10a} at similar luminosity ($L_\mathrm{bol}<10^{47}$ erg s$^{-1}$).
 Although the current sample size is small, the properties of low-luminosity quasars are in line with the recent studies of more luminous quasars at $z>5.8$ \citep{Mazzucchelli17, Shen19}.
As constant sub-Eddington accretion cannot make a $M_\mathrm{BH}\sim10^9M_\odot$ SMBH by $z\sim6$ from stellar seed black holes,
those sub-Eddington SMBHs are likely in a quiescent phase after intense mass growth, perhaps through intermittent super-Eddington phase.
 
We will continue to address the global distribution of black hole mass and Eddington ratio at $z\sim6-7$ at low luminosities  as our sample continues to grow.
 We are continuing the near-infrared follow-up observation of the SHELLQs quasars with Subaru/MOIRCS as the HSC-SSP survey proceeds and new quasars are identified.
The black hole mass measurements are combined with our host galaxy measurements with ALMA, in which we explore a less-biased view of the build up of the SMBH-galaxy co-evolution \citep{Izumi18, Izumi19}.

\acknowledgments

We thank the Gemini North and VLT staffs to execute our programs.
We are grateful to Y. Shen and K. Inayoshi for useful discussions on this work and to M. Vestergaard for kindly providing us the iron emission line templates.

This work is based on data collected at the Subaru Telescope and retrieved from the HSC data archive system, which is operated by Subaru Telescope and Astronomy Data Center at National Astronomical Observatory of Japan.

The Hyper Suprime-Cam (HSC) collaboration includes the astronomical communities of Japan and Taiwan, and Princeton University.  The HSC instrumentation and software were developed by the National Astronomical Observatory of Japan (NAOJ), the Kavli Institute for the Physics and Mathematics of the Universe (Kavli IPMU), the University of Tokyo, the High Energy Accelerator Research Organization (KEK), the Academia Sinica Institute for Astronomy and Astrophysics in Taiwan (ASIAA), and Princeton University.  Funding was contributed by the FIRST program from Japanese Cabinet Office, the Ministry of Education, Culture, Sports, Science and Technology (MEXT), the Japan Society for the Promotion of Science (JSPS),  Japan Science and Technology Agency  (JST),  the Toray Science  Foundation, NAOJ, Kavli IPMU, KEK, ASIAA,  and Princeton University.

The Pan-STARRS1 Surveys (PS1) have been made possible through contributions of the Institute for Astronomy, the University of Hawaii, the Pan-STARRS Project Office, the Max-Planck Society and its participating institutes, the Max Planck Institute for Astronomy, Heidelberg and the Max Planck Institute for Extraterrestrial Physics, Garching, The Johns Hopkins University, Durham University, the University of Edinburgh, Queen's University Belfast, the Harvard-Smithsonian Center for Astrophysics, the Las Cumbres Observatory Global Telescope Network Incorporated, the National Central University of Taiwan, the Space Telescope Science Institute, the National Aeronautics and Space Administration under Grant No. NNX08AR22G issued through the Planetary Science Division of the NASA Science Mission Directorate, the National Science Foundation under Grant No. AST-1238877, the University of Maryland, and Eotvos Lorand University (ELTE).
 
This paper makes use of software developed for the Large Synoptic Survey Telescope. We thank the LSST Project for making their code available as free software at http://dm.lsst.org.

This publication has made use of data from the VIKING survey from VISTA at the ESO Paranal Observatory, programme ID 179.A-2004. Data processing has been contributed by the VISTA Data Flow System at CASU, Cambridge and WFAU, Edinburgh.

This work was supported by JSPS KAKENHI Grant Numbers JP15J02115 (MO), JP15H03645 (NK), JP17H04830 (YM), JP15K05030 (MI), JP16H03958, JP17H01114 (TN),  JP18J01050 (YT), and the Mitsubishi Foundation Grant No. 30140 (YM).

\vspace{5mm}
\facilities{Subaru(HSC), VLT(X-Shooter), Gemini(GNIRS)}


\software{astropy \citep{Astropy},  
          SExtractor \citep{SExtractor}
          }

\bibliographystyle{aasjournal}
\bibliography{ref_SHELLQsNIR}

\end{document}